\documentclass[aps,prl,twocolumn,preprintnumbers,amsmath,amssymb,floatfix,longbibliography,superscriptaddress,10pt]{revtex4-2}

\usepackage{xcolor}
\usepackage{dsfont}
\usepackage{comment}
\usepackage{import}
\usepackage[colorlinks=true,allcolors=blue,urlcolor=blue]{hyperref}
\usepackage[sort&compress]{natbib}

\usepackage{amsthm}

\usepackage{amssymb} 
\usepackage[version=4]{mhchem} 

\usepackage{empheq} 
\usepackage{enumerate} 
\usepackage{subfiles} 

\usepackage[super]{nth}
\usepackage{comment} 
\usepackage{stackrel}
\usepackage{cancel} 

\usepackage{empheq} 

\newcommand{\innera}[2]{\left\langle#1,#2\right\rangle}

\newcommand{\expect}[1]{\left\langle#1\right\rangle}

\renewcommand{\vec}[1]{\mathbf{#1}}

\newcommand{\vecg}[1]{\boldsymbol{#1}}
\newcommand{\vk}{\mathbf{k}}
\newcommand{\vkp}{\mathbf{k}^\prime}

\newcommand{\vq}{\mathbf{q}}

\newcommand{\vz}{\mathbf{z}}

\renewcommand{\vr}{\mathbf{r}}
\newcommand{\vrp}{\mathbf{r}^\prime}

\newcommand{\vn}{\mathbf{n}}

\newcommand{\va}{\vec{a}}
\newcommand{\vb}{\vec{b}}

\newcommand{\vv}{\vec{v}}

\newcommand{\FS}{\text{FS}}

\renewcommand{\d}{\text{d}}

\newcommand{\partiald}[2]{\frac{\partial #1}{\partial #2}}

\newcommand{\expb}[1]{\exp \left[#1\right]}

\DeclareMathOperator{\tr}{tr}
\DeclareMathOperator{\Tr}{Tr}

\begin{document}
\title{Strain-induced Berry phase in chiral superconductors}

\author{Canon Sun}
\email{canon@ualberta.ca}
\affiliation{Department of Physics, University of Alberta, Edmonton, Alberta T6G 2E1, Canada}
\affiliation{Theoretical Physics Institute \& Quantum Horizons Alberta, University of Alberta, Edmonton, Alberta T6G 2E1, Canada}

\author{Marcel Franz}
 \affiliation{Department of Physics and Astronomy \& Stewart Blusson Quantum Matter Institute, University of British Columbia, Vancouver BC, Canada, V6T 1Z4}

\author{Joseph Maciejko}
\email{maciejko@ualberta.ca}
\affiliation{Department of Physics, University of Alberta, Edmonton, Alberta T6G 2E1, Canada}
\affiliation{Theoretical Physics Institute \& Quantum Horizons Alberta, University of Alberta, Edmonton, Alberta T6G 2E1, Canada}

\begin{abstract}
    We study the topology of the order parameter in the intermediate phase between the superconducting and time-reversal symmetry breaking transitions of a $p_x+ip_y$ superconductor under strain. The application of in-plane strain reduces the underlying crystal symmetry and lifts the degeneracy of the critical temperature between the $p_x$ and $p_y$ orbitals, resulting in a Dirac cone structure in the thermodynamic phase diagram. When the strain is varied adiabatically along a closed path enclosing the Dirac cone, the order parameter acquires a Berry phase of $\pi$, which originates from a \emph{half} rotation of the superconducting gap function. This half rotation leaves a topological signature in the superfluid stiffness tensor, making it directly observable through the geometry of vortices and the upper critical field.
\end{abstract}
\maketitle

\makeatother
\emph{Introduction}---A central objective of condensed matter physics is the characterization of unconventional superconducting states~\cite{Sigrist1991Phenomenological}. Among unconventional superconductors, a particularly rich class are superconductors whose order parameters have multiple components. They can exhibit behavior not present in single-component superconductors, such as spontaneous time-reversal symmetry breaking~\cite{kallin2016}, nematic order~\cite{fu2010,Fu2014Odd-parity}, the formation of half-quantum vortices~\cite{Ivanov2001Non-Abelian, Almoalem2024Observation}, and Leggett modes~\cite{Leggett1966Number}. The multicomponent nature may arise due to symmetry, where the order parameter transforms under a higher-dimensional irreducible representation (irrep) of the symmetry group. The most notable example is superfluid ${}^3\ce{He}$, a spin triplet, $p$-wave paired superfluid with nine complex components for its order parameter. In solid-state systems, there is evidence that $\ce{UPt_3}$~\cite{Joynt2002The}, UTe$_2$~\cite{Paglione2021} and doped $\ce{Bi_2Se_3}$~\cite{Yonezawa2019Nematic} host order parameters that transform in two-dimensional representations of their respective point groups. The multicomponent nature need not originate from symmetry, as in multiband superconductors~\cite{Kondo2002Theory}, such as $\ce{MgB_2}$~\cite{Nagamatsu2001Superconductivity}, iron-based superconductors~\cite{Stewart2011Superconductivity}, and the Dirac semimetal $\ce{Cd_3As_2}$~\cite{Cuozzo2024Leggett}.  In $\ce{Sr_2RuO4}$, there is evidence of a multicomponent order parameter, although its nature remains unresolved~\cite{Ghosh2021Thermodynamic, Benhabib2021Ultrasound, Kivelson2020Proposal}.

A characteristic feature of symmetry-protected multicomponent superconductors is the emergence of split superconducting transitions under symmetry-breaking perturbations~\cite{Sigrist1991Phenomenological}. Because of symmetry, in the absence of perturbation, all the components of the order parameter have the same critical temperature. When a symmetry-breaking perturbation is applied, however, this degeneracy is lifted, and components not related by symmetry condense at distinct critical temperatures. For example, in $\ce{UPt_3}$, a double transition is observed, which is attributed to the splitting of a two-dimensional irrep through coupling to magnetic order~\cite{Joynt2002The}. In superfluid $\ce{^3He}$, the degeneracy can be lifted and engineered through nanoscale confinement~\cite{li1988, Sun2023Superfluid, levitin2013}. In the case of $\ce{Sr_2RuO4}$, there is evidence of split superconducting and time-reversal symmetry breaking transitions under the application of strain~\cite{Grinenko2021Split,Grinenko2021UnSplit,grinenko2023}, although it remains controversial~\cite{Mueller2023Constraints,jerzembeck2024}. Thus, the occurrence of split transitions under symmetry-breaking fields is one of the key signatures of symmetry-protected multicomponent superconductors.

\begin{figure}[t!]
\centering
\includegraphics[width=\columnwidth]{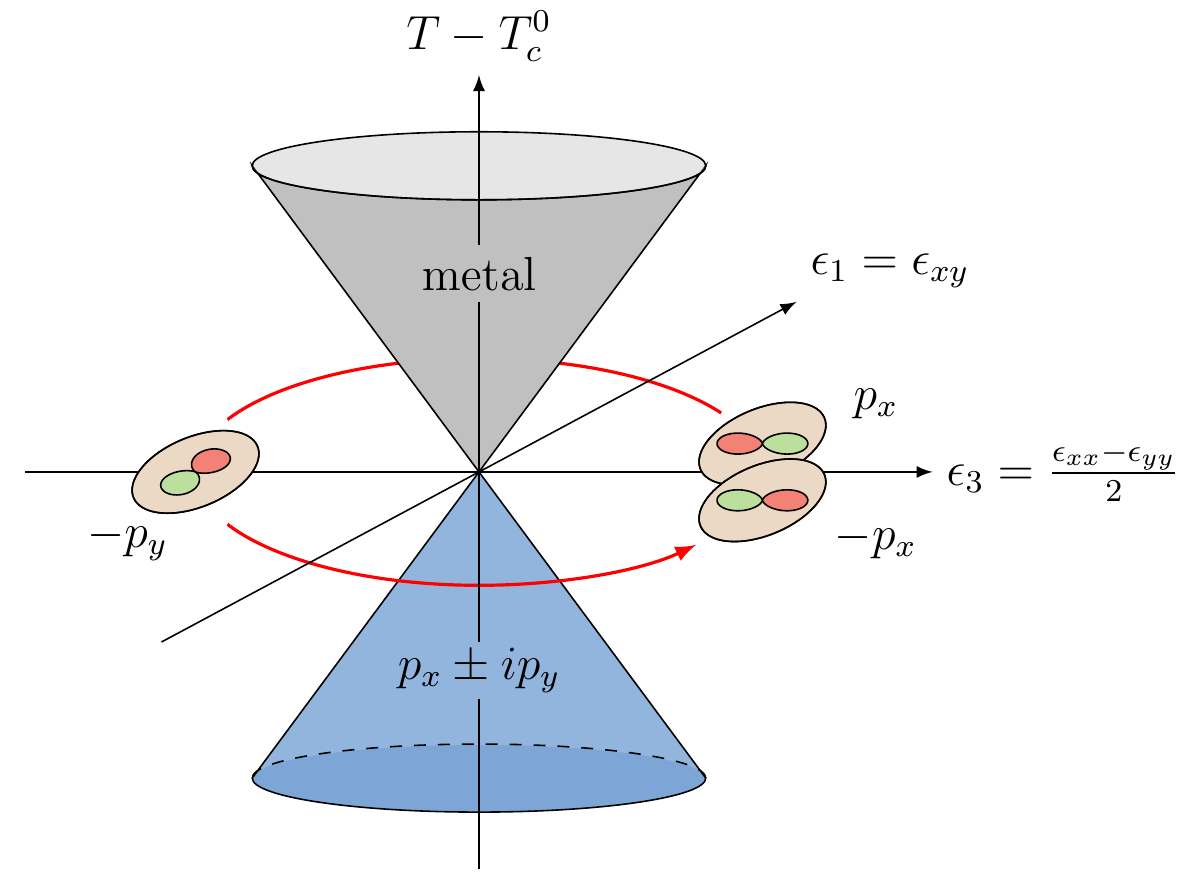}
\caption{\label{fig: schematic} Theoretical phase diagram of a $p_x\pm i p_y$ superconductor under strain. The vertical axis denotes temperature $T$ relative to the critical temperature $T_c^0$ in the absence of strain, and the horizontal axes $\epsilon_3\equiv (\epsilon_{xx}-\epsilon_{yy})/2$ and $\epsilon_1\equiv \epsilon_{xy}$ represent components of the in-plane strain tensor associated with tension and shear, respectively. The phase diagram exhibits a Dirac cone structure: the region inside the upper cone (silver) corresponds to the metallic phase, while the lower cone (blue) corresponds to the $p_x\pm ip_y$ superconductor. When the strain is varied adiabatically in a loop that encloses the Dirac cone, the gap function rotates by $\pi$.}
\end{figure}

In this Letter, we study the topology of the phase diagram of a multicomponent superconductor under symmetry-breaking perturbations. Focusing on the case of a $p_x+ip_y$ superconductor on a tetragonal lattice, when tensile and shear strain are applied, the broken symmetry splits the phase transition into two: first, a superconducting transition, followed by a time-reversal symmetry breaking transition to the chiral $p_x + i p_y$ state. The resulting split transitions form a Dirac cone structure in the phase diagram (Fig.~\ref{fig: schematic}). When the strain is varied adiabatically along a closed path enclosing this Dirac cone, the order parameter acquires a Berry phase of $\pi$, revealing the topology of the phase diagram. This Berry phase arises from a half rotation of the superconducting gap function, which, unlike the Berry phase itself, can be detected through non-phase-sensitive measurements. In particular, it manifests as a nontrivial winding in the symmetry-adapted components of the superfluid stiffness tensor, which is reflected in the real-space vortex profile and the upper critical field.

\emph{Topology of split transitions under strain}---Consider a quasi-two-dimensional $p_x+ip_y$ superconductor on a tetragonal lattice with $D_{4h}$ point group, with electrons treated as spinless for simplicity.  The system is described by the Hamiltonian
\begin{equation}\label{eq: Ham}
    \hat{H}=\hat{H}_0+\hat{H}_{\text{int}},
\end{equation}
where
\begin{align}
    \hat{H}_0 &= \sum_\vk \hat{c}^\dagger_\vk \xi_\vk \hat{c}_\vk,\\
    \hat{H}_{\mathrm{int}}&=-\frac{1}{2\mathcal{V}}\sum_{\vk\vkp}V_{\vk\vkp}\hat{c}^\dagger_{\vk}\hat{c}^\dagger_{-\vk}\hat{c}_{-\vkp}\hat{c}_{\vkp}.
\end{align}
Here, $\hat{c}^\dagger_\vk$ is the creation operator for an electron with wavevector $\vk$ and $\mathcal{V}$ is the system volume. The first term, $\hat{H}_0$, is the band Hamiltonian with single-particle dispersion
$
\xi_\vk = \sum_{\alpha=x,y}\hbar^2k_\alpha^2/(2m_0)-2t_c\cos (k_zc)-E_{0},
$
 where $m_0$ is the in-plane  effective mass, $t_c$ is the interlayer hopping amplitude, $c$ is the lattice constant along the $c$ axis, and $E_0$ is an energy shift. We suppose $t_c\ll E_{0}$, so that the Fermi surface is approximately cylindrical with radius $k_{\mathrm{F}}=\sqrt{2m_0E_{0}}/\hbar$. The interaction Hamiltonian, $\hat{H}_{\text{int}}$, describes the scattering of Cooper pairs with matrix element $V_{\vk\vkp}= V\sum_{\alpha}  \phi_\alpha(\vk)\phi_\alpha^*(\vkp)$, where $V>0$ is the interaction strength. The form factors $\phi_x(\vk) = \sqrt{2}k_x/k_{\text{F}}$ and $\phi_y(\vk) = \sqrt{2} k_y/k_{\text{F}}$ correspond to the $p_x$ and $p_y$ orbitals, which transform in the $E_u$ representation of $D_{4h}$.

We subject the system to in-plane strain, which alters the Fermi surface geometry and reduces the point group to $C_{2h}$. The strain $\epsilon$ distorts the lattice, altering both the direction and amplitude of electronic hopping. As shown in Supplemental Material (SM) Sec.~I, these effects modify the dispersion to $\xi_\vk(\epsilon) = \hbar^2 \sum_{\alpha\beta}k_\alpha M_{\alpha\beta}^{-1}(\epsilon) k_\beta/2 - 2t_c \cos(k_z c) - E_0(\epsilon) $, where the reciprocal effective mass tensor is given by $M^{-1}(\epsilon) = m_0^{-1}I_2 + \delta M^{-1}(\epsilon)$, with
\begin{equation}
    \frac{\delta M^{-1}(\epsilon)}{m_0^{-1}}\simeq -(C-2)\epsilon_0 I_2 - (C-2) \epsilon_3\sigma_3 + \epsilon_1 \sigma_1.
\end{equation}
Here, $I_2$ is the $2\times 2$ identity matrix, $\sigma_{1,2,3}$ are the Pauli matrices, $C$ is a material-dependent parameter of order one that characterizes how the hopping amplitudes decay with distance, and we assume $C>2$ without loss of generality. The coefficients $\epsilon_0\equiv(\epsilon_{xx}+\epsilon_{yy})/2$, $\epsilon_3\equiv (\epsilon_{xx}-\epsilon_{yy})/2$, and $\epsilon_1\equiv \epsilon_{xy}$ are symmetry-adapted components of the strain tensor $\epsilon_{\alpha\beta}$ that transform in the $A_{1g}$, $B_{1g}$, and $B_{2g}$ representations of $D_{4h}$, respectively. Furthermore, under strain, the band minimum is shifted to $-E_0(\epsilon)=-E_0(1-C \epsilon_0)$. For simplicity, we assume the interaction Hamiltonian is unchanged under the application of strain.

As the symmetry group is reduced under the application of strain~\cite{hicks2025}, the Ginzburg-Landau free energy describing the superconducting phase transition is modified accordingly. Associated with each partial wave channel $\phi_\alpha(\vk)$ is an order parameter field $\Delta_\alpha(\vr)$ that describes the condensation of Cooper pairs in that orbital. Expanding the free energy in powers of the order parameter fields, we obtain, to fourth order,
\begin{align}\label{eq: free energy}
    F= \int \d^3 r &\left(\sum_{\alpha\beta}\Delta^*_\alpha A_{\alpha\beta}\Delta_\beta + \sum_{ij,\alpha\beta}K_{ij,\alpha\beta}\partial_i \Delta^*_\alpha\partial_j \Delta_\beta\right. \nonumber\\
    &+ \left.\frac{1}{2}\sum_{\alpha\beta\gamma\delta}B_{\alpha\beta\gamma\delta} \Delta^*_\alpha\Delta^*_\beta\Delta_\gamma\Delta_\delta \right),
\end{align}
where $i,j=x,y,z$. The derivation of the free energy is presented in SM Sec.~II. The quadratic coefficients to linear order in strain are given by
\begin{equation}\label{A}
    A_{\alpha\beta}\simeq \frac{\delta_{\alpha\beta}}{2}\left[N_0(0) t - \frac{(C-2)\epsilon_0}{V} \right]+ \frac{1}{2V}\frac{\delta M^{-1}_{\alpha\beta}}{m_0^{-1}},
\end{equation}
where $N_0(0)=m_0/(2\pi  \hbar^2 c)$ is the density of states at the Fermi level in the absence of strain and $t= (T-T_c^0)/T_c^0$ is the reduced temperature. Here, $T_c^0= 2\Lambda e^\gamma e^{-1/[N_0(0)V]}/\pi$ is the critical temperature in the absence of strain, where $\Lambda$ is the energy cutoff and $\gamma$ is the Euler–Mascheroni constant.  To zeroth order in strain, the coefficients for the gradient and quartic terms are given by
\begin{align}
    K_{ij,\alpha\beta}&=\frac{B_0}{2}\hbar^2\langle v_iv_j\phi_\alpha^*\phi_\beta\rangle_0,\\
B_{\alpha\beta\gamma\delta}&=B_0\langle\phi_\alpha^*\phi_\beta^*\phi_\gamma\phi_\delta\rangle_0,
\end{align}
where $v_i(\vk)= \partial_{k_i} \xi_\vk(0)/\hbar$ is the Fermi velocity in the absence of strain and $B_0 = 7\zeta(3)N_0(0)/(4\pi k_BT_c^0)^2$. Here, $\expect{f}_0 \equiv \sum_\vk \delta(\xi_\vk(0)) f(\vk)/\sum_\vk \delta(\xi_\vk(0))$ denotes averaging over the (unstrained) Fermi surface.

While, in the absence of strain, the partial wave channels $\phi_x(\vk)$ and $\phi_y(\vk)$ share the same critical temperature, this degeneracy is lifted under the application of strain~\cite{Yu2019Effect}. On a mean-field level, the superconducting phase transition occurs when one of the eigenvalues of the matrix $A$ in Eq.~(\ref{A}) vanishes. In the unstrained case, the two eigenvalues of $A$ are degenerate by tetragonal symmetry, causing the partial wave channels $\phi_x(\vk)$ and $\phi_y(\vk)$ to share the same critical temperature. The thermodynamic ground state is therefore a linear combination of the two orbitals, with the specific combination determined by minimizing the full Landau free energy. For the mean-field values of the quartic coefficients, the system condenses into the fully-gapped, chiral $p_x\pm ip_y$ state. This degeneracy is lifted upon the application of strain, giving rise to the Dirac cone structure illustrated in Fig.~\ref{fig: schematic}. The system first transitions to the superconducting phase when the lowest eigenvalue of $A$ becomes negative, and, when the temperature is further lowered, there is an additional time-reversal symmetry breaking transition to the chiral $p$-wave state.

The order parameter possesses a nontrivial topology in the region of the phase diagram between the superconducting and time-reversal breaking transitions.  In this region, the order parameter is proportional to the eigenvector of $A$, or equivalently $\delta M^{-1}$, associated with the lowest eigenvalue. Let us parameterize $\cos\varphi = (C-2)\epsilon_3/\sqrt{[(C-2)\epsilon_3]^2+\epsilon_1^2}$ and $\sin\varphi = \epsilon_1/\sqrt{[(C-2)\epsilon_3]^2+\epsilon_1^2}$, so that $\varphi$ characterizes the azimuthal angle on the $\epsilon_3$-$\epsilon_1$ plane. The mean-field order parameter has the form $\vecg{\Delta}=(\Delta_x,\Delta_y)^T = \psi_0 \hat{\vecg{\Delta}}_-$, where $\hat{\vecg{\Delta}}_-=[\cos(\varphi/2),-\sin(\varphi/2)]^T$ is the normalized eigenvector of $A$ with the lowest eigenvalue. The amplitude satisfies $|\psi_0|= \sqrt{-a_-/b_-}$, where $a_-$ is the lowest eigenvalue of $A$ and $b_-= \sum_{\alpha\beta\gamma\delta}B_{\alpha\beta\gamma\delta}\hat{\Delta}^*_{-,\alpha}\hat{\Delta}^*_{-,\beta}\hat{\Delta}_{-,\gamma}\hat{\Delta}_{-,\delta}$ is the projected quartic coefficient. Because of the $\varphi/2$ dependence of $\hat{\vecg{\Delta}}_-$, the order parameter acquires a Berry phase of $\pi$ when the strain is varied adiabatically along a closed path that winds around the Dirac cone an odd number of times. Conversely, if the path winds around it an even number of times, no Berry phase is acquired. This $\pi$ Berry phase can be detected via the Josephson effect~\cite{Sun2025Topological}.

\begin{figure}[t!]
\centering
\includegraphics[width=0.78\columnwidth]{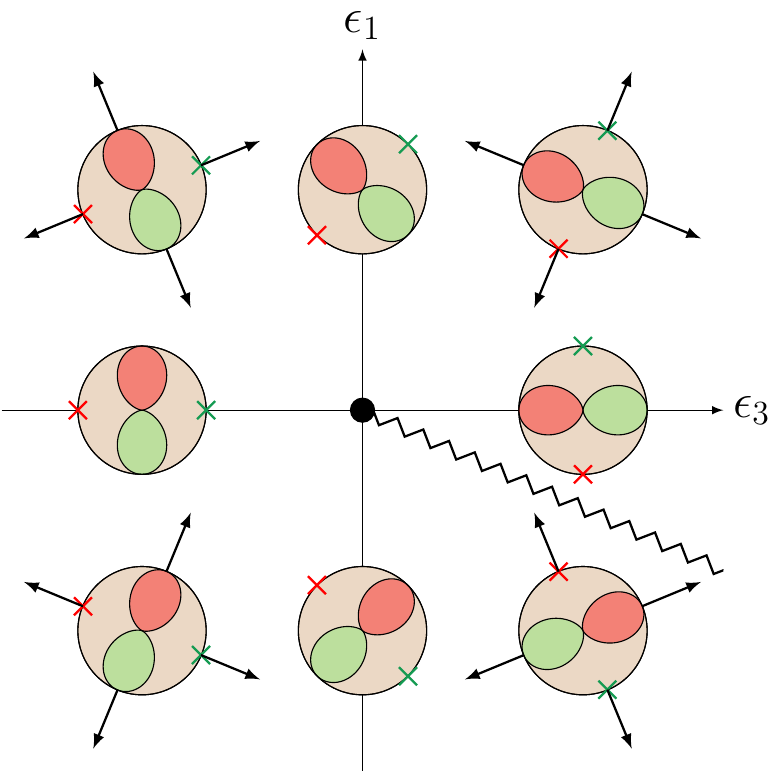}
\caption{\label{fig: orbitals} Evolution of the gap function as the strains $\epsilon_3$ and $\epsilon_1$ are varied adiabatically. The gap function is depicted as a $p$ orbital inside a constant $k_z$ slice of the Fermi sea (peach disc), and its nodes are indicated by crosses. The two lobes and nodes of the gap are distinguished by their colors. The directions of the Fermi velocity at the gap maxima and nodes are indicated with arrows. When the strain is varied adiabatically in a loop that encloses the Dirac point at the origin (black dot), the gap function only performs half a rotation, indicating the presence of a branch cut in the phase diagram (wavy line).}
\end{figure}

The $\pi$ Berry phase signifies that the superconducting gap function has rotated about the $k_z$ axis by $\pi$ at the end of the adiabatic evolution. The mean-field gap function is given by $\Delta_\vk \equiv  \sum_\alpha \Delta_\alpha(\varphi) \phi_\alpha(\vk)=\sqrt{2}\psi_0\cos(\theta_\vk + \varphi/2) $, where $\theta_\vk$ is the polar angle in the $k_x$-$k_y$ plane. As the strain parameter $\varphi$ is varied adiabatically through $2\pi$, the gap function rotates in momentum space, but at only half the speed (Fig.~\ref{fig: orbitals}). As a result, after completing a loop that encloses the Dirac cone, the gap function only performs half a rotation. In particular, compared to the initial configuration, the two lobes of the $p$ orbital, as well as the two nodes, have swapped positions. By contrast, after traversing in a contractible loop, i.e., $\varphi$ returns back to itself, the lobes and nodes return back to their original locations.

\emph{Topology of the superfluid stiffness tensor}---By tracking the orientation of the gap function throughout the adiabatic process, achievable through non-phase-sensitive measurements as we argue below, one could detect whether the gap function has performed a half rotation. By leveraging the anisotropy of the gap function, the topology of the path taken in strain space manifests in the superfluid stiffness tensor, which determines the shape of Abrikosov vortices as well as the upper critical field.

The low-energy behavior of the superconductor between the split transitions can be described by a single-component Ginzburg-Landau theory. As the low-energy excitations consist only of spatial variations in the condensate amplitude, we make the ansatz $\vecg{\Delta}(\vr) = \psi(\vr) \hat{\vecg{\Delta}}_-$, and the Ginzburg-Landau free energy reduces to
\begin{equation}\label{eq: projected F}
    F = \int \d^3 r \left(a_- |\psi|^2 + \sum_{ij}K_{ij}\partial_i \psi^*\partial_j \psi + \frac{b_-}{2} |\psi|^4  \right),
\end{equation}
where the projected superfluid stiffness tensor is given by~\footnote{Here, the tensor $K_{ij}$ is the stiffness associated with spatial variations of the amplitude $\psi(\vr)$. This is related to the more standard definition of the superfluid stiffness $D_{ij}=K_{ij}|\psi|^2$~\cite{Kreidel2024Measuring}, which measures the energetic cost of phase fluctuations.}
\begin{align}\label{eq: K_ij}
    K_{ij}&\equiv \sum_{\alpha\beta}K_{ij,\alpha\beta}\hat{\Delta}^*_{-,\alpha}\hat{\Delta}_{-,\beta}=\frac{B_0}{2} \hbar^2 \expect{v_i v_j|g|^2}_{0},
\end{align}
with normalized gap function $g(\vk) = \sum_\alpha \hat{\Delta}_{-,\alpha}(\varphi)\phi_\alpha(\vk)$. Note that, to lowest order, the strain enters the superfluid stiffness tensor through $g(\vk)$, the angular dependence of the gap function, which depends only on $\varphi$ and not the strain amplitude. In other words, for $K_{ij}$, the strain acts as a symmetry-breaking field, functioning only to select a particular $g(\vk)$, analogous to nematic superconductors. As expressed in Eq.~\eqref{eq: projected F}, the Ginzburg–Landau free energy describes an anisotropic superconductor characterized by a single complex scalar field $\psi$.

\begin{figure}[t!]
\centering
\includegraphics[width=0.98\columnwidth]{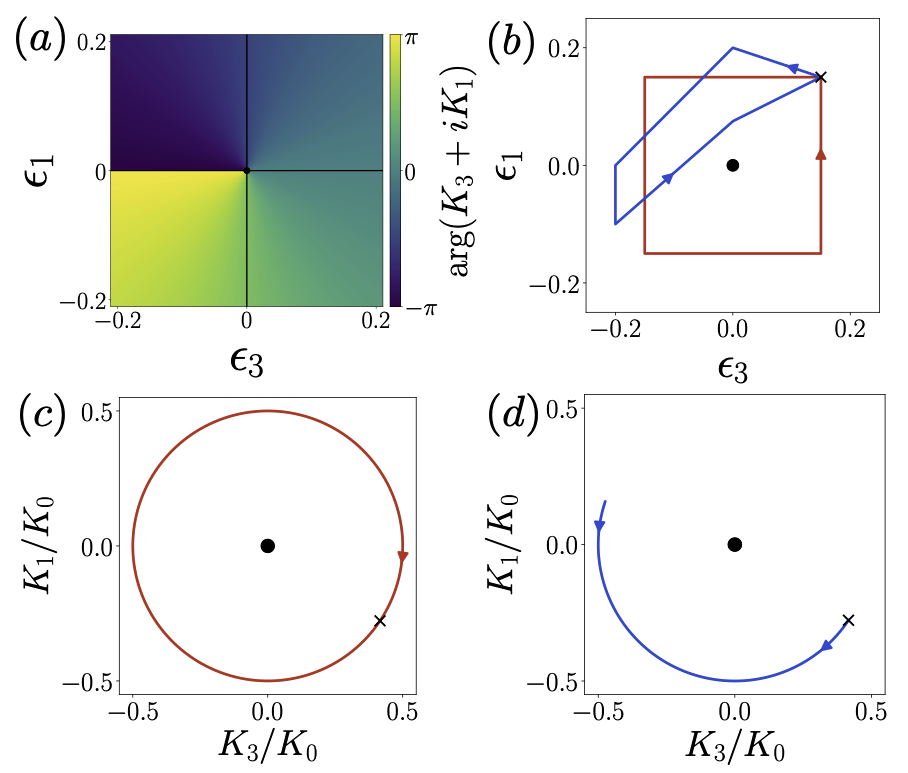}
\caption{\label{fig: stiffness} (a) Phase of $K_3 + iK_1$ on the $\epsilon_3$–$\epsilon_1$ plane, showing $2\pi$ winding around the origin. (b) Two closed paths in strain space: a loop that winds once around the origin (red) and a contractible loop (blue). Their corresponding trajectories on the $K_3$–$K_1$ plane are shown in (c) and (d), respectively. The red path winds once around the origin, whereas the blue path backtracks to its starting position and is topologically trivial. The black dot marks the origin, and the cross indicates the starting point of the loops. Parameter values are $\Lambda = 10$, $N_0(0)V = 0.3$, $C = 3.5$, and $\epsilon_0=0$.}
\end{figure}

The orientation of the gap function can be determined from the following two symmetry-adapted components of the superfluid stiffness tensor:
\begin{align}
    K_3&\equiv \frac{K_{xx}-K_{yy}}{2},&K_1&\equiv K_{xy}.
\end{align}
To understand how the orientation of the gap can be inferred from $K_3$ and $K_1$, suppose, for example, $\epsilon_3, \epsilon_1 > 0$. From the Fermi velocity vectors (black arrows) in the first quadrant of Fig.~\ref{fig: orbitals}, we observe that $v_x^2 - v_y^2 > 0$ and $v_x v_y < 0$ around the gap maxima, whereas they have the opposite signs around the nodes. Because the Fermi surface average in Eq.~(\ref{eq: K_ij}) is weighted by the gap function $|g(\vk)|^2$, contributions from the lobes would dominate over those from the nodes. Therefore, $K_3 > 0$ and $K_1 < 0$ in that quadrant. Repeating the same reasoning for the other quadrants of Fig.~\ref{fig: orbitals}, we conclude that $K_3$ has the same sign as $\epsilon_3$ while $K_1$ has the opposite sign as $\epsilon_1$. Indeed, the Fermi surface average yields
\begin{align}\label{eq: K3 and K1}
    K_3&=\frac{K_0}{2}\cos\varphi,&K_1&=-\frac{K_0}{2}\sin\varphi,
\end{align}
where $K_0\equiv (K_{xx}+K_{yy})/2= B_0\hbar^4k_{\text{F}}^2/(8m_0^2)$. The phase of the complex combination $K_3+i K_1$ winds by $2\pi$ around the origin of the $\epsilon_3$-$\epsilon_1$ plane [Fig.~\ref{fig: stiffness}(a)], and the ratio $K_1/K_3$ directly encodes the orientation of the gap nodes. Note that $K_3$ and $K_1$ are ill defined at the origin, which reflects the fact that the strain amplitude acts as a symmetry-breaking field, and gives different angular dependence for the gap function $g(\vk)$ when approached from different angles $\varphi$.

The rotation of the gap function along a closed loop in strain space can be tracked by analyzing the topology of the path traced out in the $K_3$-$K_1$ plane. According to Eq.~\eqref{eq: K3 and K1}, when the strain is varied in a way that encircles the origin once, i.e., $\varphi\mapsto \varphi+2\pi$, the corresponding path in the $K_3$-$K_1$ plane likewise winds once about its origin. On the other hand, if $\varphi$ follows a contractible loop in the $\epsilon_3$-$\epsilon_1$ plane, the resulting trajectory in the $K_3$-$K_1$ plane is also contractible. To verify these predictions, we compute $K_3$ and $K_1$ for two topologically inequivalent loops in strain space by numerically minimizing the Landau free energy. Fig.~\ref{fig: stiffness}(b) shows two closed paths in strain space: a loop that winds once around the Dirac point (red) and a contractible loop (blue). Figs.~\ref{fig: stiffness}(c) and (d) display their corresponding trajectories in the $K_3$–$K_1$ plane, demonstrating that the red path winds around the origin, whereas the blue path does not. Therefore, the topology of the path in the $K_3$-$K_1$ plane indicates whether the gap function has performed a half rotation.

\emph{Shape of Abrikosov vortices}---The superfluid stiffness tensor determines the shape of Abrikosov vortices. Consider an isolated vortex along the $z$-axis at the origin. To solve the Ginzburg-Landau equation, we first perform the transformation $\vrp = \xi^{-1}\vr$, where $\xi=\sqrt{K}/\sqrt{|a_-|}$ is the healing length tensor~\footnote{The matrix square root $\sqrt{K}$ exists because the superfluid stiffness tensor $K$ is positive semidefinite.}, so that the equation becomes isotropic. The vortex solution is $\psi(\vr) = \psi_0 f(\rho^\prime) e^{i \theta^\prime} $, where $\rho^\prime $ and $\theta^\prime$ are the radial and azimuthal coordinates of $\vrp$ in cylindrical coordinates. Here, $f(\rho^\prime)$ satisfies the differential equation $[\partial_{\rho^\prime}(\rho^\prime \partial_{\rho^\prime}f)]/\rho^\prime- (1/\rho^{\prime 2}-1)f- f^3=0$ with boundary conditions $f(0)=0$ and $f(\rho^\prime\to\infty)=1$. To study the geometry of the vortex, we consider level sets of the function $f(\rho^\prime)$, or equivalently $\rho^\prime$, which satisfy
\begin{equation}
    \sum_{\alpha\beta}r_\alpha \xi^{-2}_{\alpha\beta}(\epsilon)r_\beta= \text{const.}
\end{equation}
In other words, the vortex has an elliptical profile in the $x$-$y$ plane, with semimajor axis $\vec{u}_+(\epsilon)=\xi_+(\epsilon)[\cos(\varphi/2), -\sin(\varphi/2)]^T$ and semiminor axis $\vec{u}_-(\epsilon)=\xi_-(\epsilon)[\sin(\varphi/2), \cos(\varphi/2)]^T$, where $\xi_{\pm}(\epsilon)=\sqrt{(K_0\pm K_0/2)/|a_-(\epsilon)| }$.

\begin{figure}[t!]
\centering
\includegraphics[width=0.98\columnwidth]{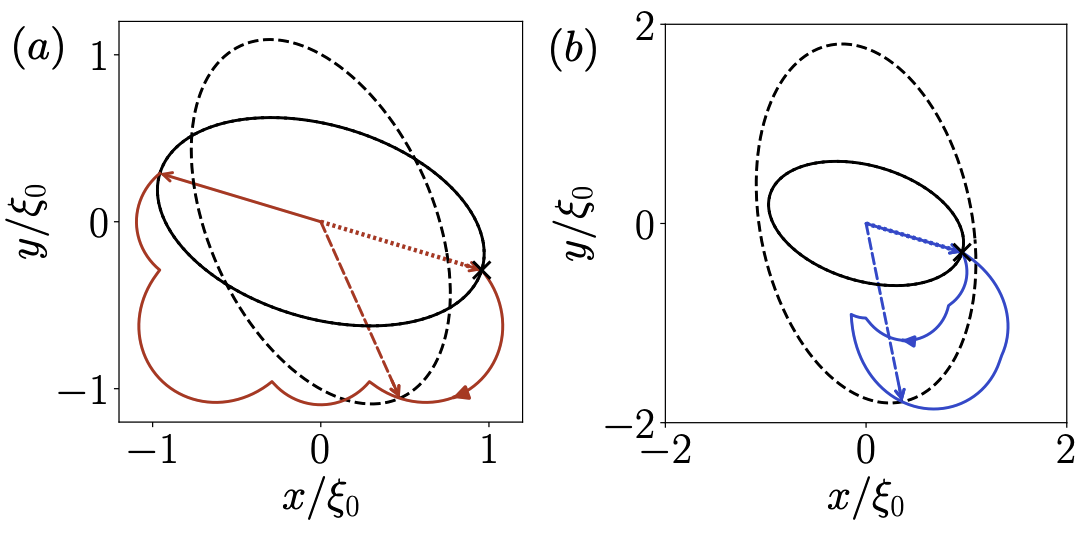}
\caption{\label{fig: vortex} Evolution of the vortex profile as strain is varied along the two paths in Fig.~\ref{fig: stiffness}(b), with (a) corresponding to the red path and (b) to the blue path. The vortex is represented by an ellipse whose semimajor axis is indicated by an arrow. The initial, intermediate, and final configurations are shown in dotted, dashed, and solid lines, respectively. Note that the initial and final ellipses overlap, and the arrows also coincide in (b). The cross indicates the initial, arbitrarily chosen orientation of the semimajor axis, and the colored curve shows how the semimajor axis evolves as the strain is varied. In (a), the semimajor axis flips orientation, whereas in (b) it returns to its original configuration. Here, $\xi_0$ denotes the initial length of the semimajor axis.}
\end{figure}

The topology of the superfluid stiffness tensor is reflected in the evolution of the vortex profile geometry (Fig.~\ref{fig: vortex}). Because $\xi_+(\epsilon) \neq \xi_-(\epsilon)$, the vectors $\vec{u}_+(\epsilon)$ and $\vec{u}_-(\epsilon)$ defining the semimajor and semiminor axes can be uniquely identified up to an overall sign. We pick an initial direction for the semimajor (or semiminor) axis, and track its evolution along the paths in Fig.~\ref{fig: stiffness}(b). When the strain parameter $\varphi$ is varied continuously from $\varphi$ to $\varphi + 2\pi$ (red path), the semimajor axis reverses its orientation at the end of the cycle [Fig.~\ref{fig: vortex}(a)]. Conversely, for the contractible path (blue), $\vec{u}_+$ returns back to itself [Fig.~\ref{fig: vortex}(b)]. Therefore, by monitoring how the shape of Abrikosov vortices evolves under strain, one can infer the presence of a Dirac point in the thermodynamic phase diagram.

\begin{figure}[t!]
\centering
\includegraphics[width=0.8\columnwidth]{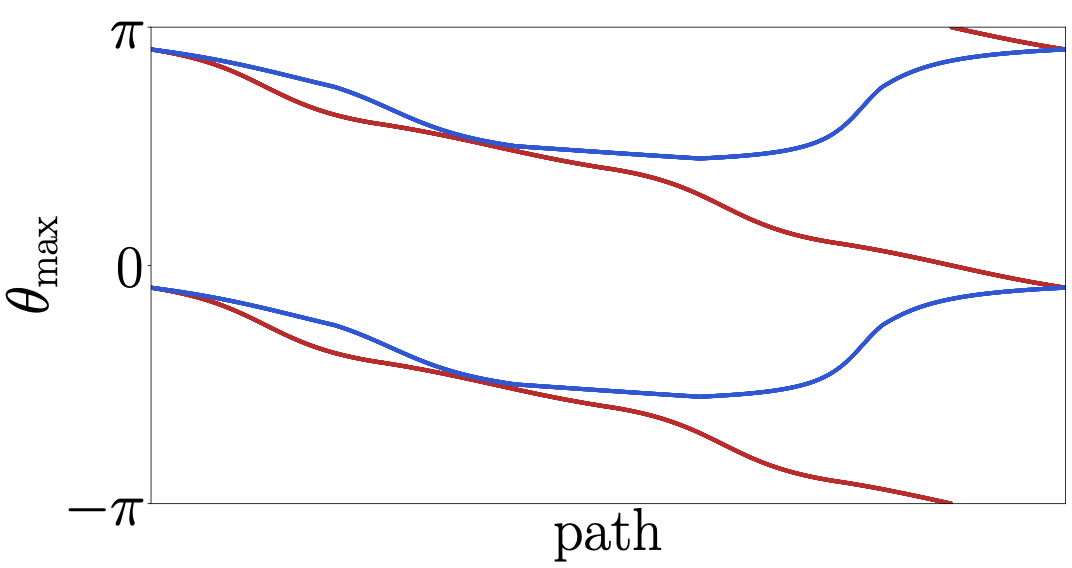}
\caption{\label{fig: critical field} Angles $\theta_{\text{max}}$ at which the in-plane upper critical field $B_{c2}$ is maximal as the strain is varied along the two paths in Fig.~\ref{fig: stiffness}(b). At the end of the evolution, the two maxima exchange positions for the red path, whereas they return to their original locations for the blue path.}
\end{figure}

\emph{Upper critical field}---The evolution of the healing-length tensor can also be probed through measurements of the upper critical field $B_{c2}$. Suppose a magnetic field is applied in the $\hat{\vec{n}}=(\cos\theta,\sin\theta,0)^T$ direction lying in the $x$-$y$ plane. The upper critical field depends on the healing lengths in the two directions perpendicular to $\hat{\vn}$ (see SM III for further details):
\begin{equation}\label{eq: Bc2}
    B_{\text{c}2}(\hat{\vec{n}})= \frac{\Phi_0}{2\pi \xi_{zz}\xi_\perp(\hat{\vec{n}})},
\end{equation}
where $\Phi_0= h/(2e)$ is the superconducting flux quantum,  $\xi_{zz}=\sqrt{K_{zz}/|a_-|}$ is the healing length in the $z$ direction, and $\xi_\perp(\hat{\vec{n}})=\sqrt{\hat{\vec{n}}_\perp^T \xi^2 \hat{\vec{n}}_\perp}$ is the healing length in the in-plane direction $\hat{\vec{n}}_\perp =\hat{\vec{z}}\times\hat{\vec{n}}$ perpendicular to $\hat{\vec{n}}$. The upper critical field is maximal when the magnetic field is aligned with a semimajor axis $\vec{u}_+$. Consequently, by monitoring the direction in which $B_{\text{c}2}$ attains its maximum, one can follow the rotation of the semimajor axis. In Fig.~\ref{fig: critical field}, we show the angles at which the critical field is maximal as the strain is varied along the two paths in Fig.~\ref{fig: stiffness}(b), revealing an inversion of the semimajor axis for the red path and lack thereof for the blue path.

Besides the superfluid stiffness tensor, which is sensitive to the location of the gap maxima, the orientation of the gap function can also be inferred from measurements that probe the nodes. If the temperature can be lowered sufficiently while maintaining the system in the intermediate phase, the thermal conductivity provides an alternative probe~\cite{Mineev1999Introduction}. In particular, the symmetry-adapted components of the thermal conductivity tensor, $\kappa_{3} \equiv (\kappa_{xx} - \kappa_{yy})/2$ and $\kappa_{1} \equiv \kappa_{xy}$, play roles analogous to $K_3$ and $K_1$ and are expected to exhibit a similar winding. However, analogous behavior may also arise in the normal state due to Fermi surface distortions induced by strain. A detailed discussion is provided in SM IV.

\emph{Conclusion}---We have studied the topology of the order parameter in the region between the superconducting and time-reversal symmetry breaking phases of a $p_x+ip_y$ superconductor under strain. We showed that, under the application of strain, the multicritical point splits and acquires a Dirac cone structure, and the order parameter exhibits a Berry phase of $\pi$ along a path encircling this cone. The topology of the order parameter is manifested in the winding of the symmetry-adapted components of the superfluid stiffness tensor, which can be probed by tracking the evolution of the real-space vortex profile and the direction along which the upper critical field is maximal. Although our analysis focused on a $p_x+ip_y$ superconductor with tetragonal symmetry, the arguments can be readily generalized to other multicomponent systems. More generally, our work provides a protocol with a topological signature that distinguishes between symmetry-protected and accidental degeneracies in the phase diagrams of unconventional superconductors.

\emph{Acknowledgements}---The authors thank Albion Arifi, Sourav Biswas, Igor
Boettcher, Alex Hickey, Davidson Noby Joseph, Jongjun M. Lee, 
Hanbang Lu, Frank Marsiglio, and Connor
Walsh for valuable discussions. C.S. acknowledges support through the Natural Sciences and Engineering Research Council of Canada (NSERC) Discovery Grant RGPAS-2020-00064 and the Pacific Institute for the Mathematical Sciences CRG PDF Fellowship Award. 
J.M. was supported by NSERC Discovery Grants RGPIN-2020-06999 and
RGPAS-2020-00064; M.F. by RGPIN-2022-03720 and ALLRP-580721-22.

\clearpage
\onecolumngrid

\section*{S\lowercase{upplemental} M\lowercase{aterial for} ``S\lowercase{train-induced} B\lowercase{erry phase in chiral superconductors}''} 

\renewcommand{\theequation}{S\arabic{equation}}
\setcounter{equation}{0}  

\section{C\lowercase{hange in the effective mass tensor under strain}}
In this section, we derive the electronic dispersion under strain based on a microscopic tight-binding model. By expanding the tight-binding Hamiltonian to linear order in strain, we find that the anisotropy in the geometry and change in hopping amplitudes modify the effective mass tensor and shift the band bottom.

Consider a collection of spinless electrons on a tetragonal lattice with $D_{4h}$ point group. They are described by the tight-binding Hamiltonian
\begin{equation}
     \hat{H}_0 = -\sum_\vr \left(t_a \hat{c}^\dagger_{\vr+\va}\hat{c}_\vr + t_b \hat{c}^\dagger_{\vr+\vb}\hat{c}_{\vr}+t_c\hat{c}^\dagger_{\vr+\vec{c}}\hat{c}_{\vr}+\text{h.c.} \right) ,
\end{equation}
where $\hat{c}^\dagger_\vr$ is the creation operator for an electron at site $\vr$. This Hamiltonian describes nearest-neighbor hopping in the $\vec{a}=\vec{a}_0\equiv a_0(1,0,0)^T$, $\vec{b}=\vec{b}_0 \equiv a_0(0,1,0)^T$, and $\vec{c} = c(0,0,1)^T$ directions, where $a_0$ and $c$ are the in-plane and out-of-plane lattice constants, respectively. We suppose the system is quasi-two-dimensional, so that the in-plane and out-of-plane hopping parameters, $t_a=t_b\equiv t_0$ and $t_c$, satisfy $t_0\gg t_c$. 

Under in-plane strain, the tight-binding Hamiltonian is affected in two ways. First, the primitive lattice vectors are modified according to 
\begin{equation}
    \begin{pmatrix}
        \vec{a}\\
        \vec{b}
    \end{pmatrix}
    =\begin{pmatrix}
        1+\epsilon_{xx}&\epsilon_{xy}\\
        \epsilon_{xy}&1+\epsilon_{yy}
    \end{pmatrix}
    \begin{pmatrix}
        \vec{a}_0\\
        \vec{b}_0
    \end{pmatrix},
\end{equation}
where $\epsilon_{\alpha\beta}$, with $\alpha,\beta=x,y$, are the components of the in-plane strain tensor. Second, the change in the distance between neighboring sites alters the hopping amplitudes, which can be modeled phenomenologically as $t_a = t_0 \expb{-C \left(|\va|/a_0 -1\right)}\simeq t_0 \left(1-C \epsilon_{xx} \right)$ and $t_b\simeq t_0(1-C\epsilon_{yy})$, where $C>0$ is a material-dependent dimensionless parameter of order one \cite{deJaun2012Space, Steppke2017Strong}. For a generic strain involving both tension and shear, the point group of the lattice is reduced from $D_{4h}$ to $C_{2h}$. 

In the momentum representation, the dispersion is given by
\begin{equation}\label{dispersion}
    \xi_\vk(\epsilon) =  \frac{\hbar^2}{2}\sum_{\alpha\beta}k_\alpha M^{-1}_{\alpha\beta}(\epsilon)k_\beta-2t_c\cos(k_z c)-E_{0}(\epsilon),
\end{equation}
where $M^{-1}(\epsilon)=m_0^{-1}I_2+\delta M^{-1}(\epsilon)$ is the reciprocal effective mass tensor, $m_0^{-1} \equiv 2t_0a_0^2/\hbar^2$ is the inverse effective mass in the unstrained system, and, to linear order in strain,
\begin{equation}
    \frac{\delta M^{-1}(\epsilon)}{m_0^{-1}}\simeq -(C-2)\epsilon_0 I_2 - (C-2) \epsilon_3\sigma_3 + \epsilon_1 \sigma_1.
\end{equation}
Here, $I_2$ is the $2\times 2$ identity matrix and $\sigma_{1,2,3}$ are the Pauli matrices. The coefficients $\epsilon_0\equiv(\epsilon_{xx}+\epsilon_{yy})/2$, $\epsilon_3\equiv (\epsilon_{xx}-\epsilon_{yy})/2$, and $\epsilon_1\equiv \epsilon_{xy}$ represent symmetry-adapted components of the strain tensor. The band minimum is at $-E_{0}(\epsilon)\simeq -E_0(1-C \epsilon_0)$, where $E_0 =  4t_0$.

\section{G\lowercase{inzburg}-L\lowercase{andau free energy}}
In this section, we derive the Ginzburg–Landau free energy given in Eq.~\eqref{eq: free energy} of the main text. We first show how to evaluate integrals over anisotropic Fermi surfaces, which commonly arise in the Bardeen-Cooper-Schrieffer (BCS) theory of superconductivity. We then derive an effective bosonic theory describing the superconducting phase transition.

Let us begin with the Hamiltonian of our system under strain:
\begin{equation}
    \hat{H}=\hat{H}_0+\hat{H}_{\text{int}},
\end{equation}
where, in the momentum representation, the band and interaction Hamiltonians are
\begin{align}
    \hat{H}_0 &= \sum_\vk \hat{c}^\dagger_\vk \xi_\vk(\epsilon) \hat{c}_\vk,\\
    \hat{H}_{\text{int}}&=-\frac{1}{2\mathcal{V}}\sum_{\vk\vkp\vq}V_{\vk\vkp}\hat{c}^\dagger_{\vk_+}\hat{c}^\dagger_{-\vk_-}\hat{c}_{-\vk^\prime_-}\hat{c}_{\vk^\prime_+}.
\end{align}
For notational compactness, we employ the short-hand notation $\vk_\pm =  \vk\pm \vq/2$. We have included a conserved center-of-mass momentum $\vq$ for the Cooper pairs to allow the order parameter to be spatially varying.

\subsection{Fermi surface harmonics}\label{SM: FS harmonics}
In the weak-coupling approximation, momentum sums are restricted to states lying within a thin shell of width $2\Lambda$ around the Fermi surface, where $\Lambda$ is a pairing scale. In the thermodynamic limit, each momentum sum is replaced by an integral, which can be decomposed into a surface integral over the Fermi surface and an integral over constant-energy surfaces parameterized by $\xi$. More explicitly, the momentum sum can be recast as
\begin{equation}
    \frac{1}{\mathcal{V}}\sum_\vk \mapsto \int \frac{\d^3k}{(2\pi)^3} \equiv\int_{-\Lambda}^\Lambda \d \xi \int \frac{\d^3k}{(2\pi)^3} \delta(\xi_\vk(\epsilon) - \xi)= \int_{-\Lambda}^\Lambda \d \xi \int_{\FS(\epsilon)} \frac{\d S_\vk}{(2\pi)^3} \frac{1}{|\nabla_\vk\xi_\vk(\epsilon)|},
\end{equation}
where $\d S_\vk$ denotes the surface element on the Fermi surface $\FS(\epsilon)$, which depends on the strain $\epsilon_{\alpha\beta}$. In this expression, the integral over the Fermi surface is weighted by a factor of $1/|\nabla_\vk\xi_\vk(\epsilon)|$, reflecting the fact that the distribution of states is not uniform over the Fermi surface. As such, for any function $f$ defined on the Fermi surface, the appropriate Fermi surface average is given by
\begin{align}
    \expect{f}_\epsilon &\left. \equiv \int \frac{\d^3k}{(2\pi)^3}\delta(\xi_\vk(\epsilon)) f(\vk)\middle/  \int \frac{\d^3k}{(2\pi)^3}\delta(\xi_\vk(\epsilon)) \right.\nonumber\\
    &= \left.\int \frac{\d S_\vk}{(2\pi)^3}\frac{1}{|\nabla_\vk\xi_\vk(\epsilon)|}f(\vk)\middle/  \int\frac{\d S_\vk}{(2\pi)^3}\frac{1}{|\nabla_\vk\xi_\vk(\epsilon)|}\right. \nonumber\\
    &= \frac{1}{N_\epsilon(0)}\int \frac{\d S_\vk}{(2\pi)^3}\frac{1}{|\nabla_\vk\xi_\vk(\epsilon)|}f(\vk),
\end{align}
where $N_\epsilon(0)$ is the density of states at the Fermi level under strain $\epsilon$. The subscript $\epsilon$ in the Fermi surface average makes explicit the strain dependence of the Fermi surface and the associated Fermi surface average.

Analogously, one could define a Fermi surface--weighted inner product for functions defined on the Fermi surface~\cite{Allen1976Fermi}:
\begin{equation}
    \innera{f}{g}_\epsilon\equiv \expect{f^*g}_\epsilon = \left.\int \frac{\d S_\vk}{(2\pi)^3}\frac{1}{|\nabla_\vk\xi_\vk(\epsilon)|}f^*(\vk)g(\vk)\middle/  \frac{\d S_\vk}{(2\pi)^3}\frac{1}{|\nabla_\vk\xi_\vk(\epsilon)|}\right. .
\end{equation}
In order to classify the possible pairing states, one would construct basis functions on the Fermi surface that transform according to irreducible representations (irreps) of the symmetry group and are orthonormal with respect to this inner product. More explicitly, let $\chi^{(\mu)}_{\alpha m}$ be a function on the Fermi surface transforming in the irrep $\mu$ of the symmetry group, with $\alpha$ labeling the multiplicity of the irrep and $m=1,\dots,d_\mu$, where $d_\mu$ is the dimension of the irrep $\mu$, denoting components that are mixed under symmetry operations. Given a fixed Fermi surface specified by the strain tensor $\epsilon$, the functions can be normalized to
\begin{equation}
    \innera{\chi^{(\mu)}_{\alpha m}}{\chi^{(\nu)}_{\beta n}}_\epsilon=\delta_{\mu\nu}\delta_{\alpha\beta}\delta_{mn}.
\end{equation}
Introduced in Ref.~\cite{Allen1976Fermi}, the polynomials $\chi^{(\mu)}_{\alpha m}$ are referred to as ``Fermi surface harmonics".

Basis functions that are orthogonal with respect to one Fermi surface need not remain orthogonal with respect to another, even if they have the same symmetry group. While orthogonality in the indices $\mu,\nu$ and $m,n$ is enforced by symmetry, no such constraint applies to the multiplicity index. In the context of our work, the basis functions $\phi_x$ and $\phi_y$ both transform under the one-dimensional $B_u$ irrep of the group $C_{2h}$. Consequently, as we show below, the Fermi surface--weighted inner product $\innera{\phi_x}{\phi_y}_\epsilon$ varies with strain. 

Let us now explicitly demonstrate how the Fermi surface average is computed for our system. Consider the integral
\begin{equation}
	I \equiv \int \frac{\d^3k}{(2\pi)^3} f(\vk),
\end{equation}
in which it is understood that the domain of integration is over a thin shell around the Fermi surface. As derived in Eq.~(\ref{dispersion}), the dispersion is given by
\begin{equation}
    \xi_\vk(\epsilon) = \frac{\hbar^2}{2}\sum_{ij}k_i M_{ij}^{-1}(\epsilon)k_j - 2 t_c\cos(k_z c) -E_0(\epsilon).
\end{equation}
We assume the system is quasi-two-dimensional, in the sense that $t_c\ll E_0$, so that, for the purposes of the integral measure and domain, the dispersion can be approximated as
\begin{equation}
    \xi_\vk(\epsilon) \simeq \frac{\hbar^2}{2}\sum_{ij}k_i M_{ij}^{-1}(\epsilon)k_j -E_0(\epsilon).
\end{equation}
The Fermi surface defined by this dispersion is a cylinder with an elliptical base. To proceed, we perform the change of coordinates
\begin{align}
	\vq &= S \vk,& S&= \begin{pmatrix}
	    \sqrt{M^{-1}}/\sqrt{m_0^{-1}}&0\\
        0&1
	\end{pmatrix},
\end{align}
where $\sqrt{M^{-1}}$ is the square root matrix of $M^{-1}$, which exists because $M^{-1}$ is positive semidefinite. This brings the dispersion to a form that is invariant under rotations about the $z$-axis,
\begin{equation}
	\tilde{\xi}_\vq(\epsilon)\equiv \xi_{S^{-1}\vq}(\epsilon) = \frac{\hbar^2(q_x^2+q_y^2)}{2m_0} - E_{0}(\epsilon),
\end{equation}
while leaving the $q_z= k_z(\epsilon)$ component unchanged. The Fermi surface in $\vq$-space is a cylinder of radius $k_{\text{F}}(\epsilon)= \sqrt{2m_0 E_{0}(\epsilon)}/\hbar$ and height $k_{\text{F}}^c \equiv 2\pi/c$, where $c$ is the lattice constant in the $z$ direction. We employ the notation that a tilde on a function denotes $\tilde{f}(\vq) \equiv f(S^{-1}\vq)$. Performing the change of coordinates,
\begin{align}
    I &= (\det S^{-1})\int \frac{\d^3q}{(2\pi)^3} \tilde{f}(\vq)\nonumber\\
    &= (\det S^{-1})\int_{-\Lambda}^\Lambda \d \xi \int \frac{\d S_\vq}{(2\pi)^3}\frac{1}{|\nabla_\vq \tilde{\xi}_\vq(\epsilon)|}\tilde{f}(\vq)\nonumber\\
    &=(\det S^{-1})N_0(0)\int_{-\Lambda}^\Lambda \d \tilde{\xi}_\vq \int_0^{k_{\text{F}}^c} \frac{\d q_z}{k_{\text{F}}^c}\int_0^{2\pi} \frac{\d\theta_\vq}{2\pi} \tilde{f}(\vq),\label{eq: I1}
\end{align}
where $\theta_\vq$, defined by $\tan\theta_\vq=q_y/q_x$, is the in-plane polar angle and $N_0(0)=k_{\text{F}}^c m_0/(4\pi^2 \hbar^2)$ is the density of states in the absence of strain.

As an application of Eq.~\eqref{eq: I1}, let us calculate the density of states at the Fermi level under strain. It is defined as
\begin{equation}
    N_\epsilon(0) = \int \frac{\d^3k}{(2\pi)^3}\delta(\xi_\vk(\epsilon)).
\end{equation}
Noticing that this is simply the case when $f(\vk) = \delta(\xi_\vk(\epsilon))$, we immediately have
\begin{equation}
    N_\epsilon(0)= (\det S^{-1}) N_0(0)\simeq N_0(0)\left[1+ (C-2)\epsilon_0 \right].
\end{equation}
Therefore, we can rewrite Eq.~\eqref{eq: I1} as
\begin{equation}\label{eq: I2}
    I \equiv \int \frac{\d^3k}{(2\pi)^3}f(\vk)= N_\epsilon(0)\int_{-\Lambda}^\Lambda \d \tilde{\xi}_\vq \int_0^{k_{\text{F}}^c} \frac{\d q_z}{k_{\text{F}}^c}\int_0^{2\pi} \frac{\d\theta_\vq}{2\pi} \tilde{f}(\vq).
\end{equation}
Using this expression, the Fermi surface average of a function $f$ can be expressed as
\begin{align}\label{eq: f average}
    \expect{f}_\epsilon = \int_0^{k_{\text{F}}^c}\frac{\d q_z}{k_{\text{F}}^c}\int_0^{2\pi} \frac{\d\theta_\vq}{2\pi}\tilde{f}(\vq).
\end{align}
Similarly, the Fermi surface--weighted inner product takes the form
\begin{align}\label{eq: fg average}
    \innera{f}{g}_\epsilon = \int_0^{k_{\text{F}}^c}\frac{\d q_z}{k_{\text{F}}^c}\int_0^{2\pi} \frac{\d\theta_\vq}{2\pi} \tilde{f}^*(\vq)\tilde{g}(\vq).
\end{align}
Eqs.~\eqref{eq: I2}, \eqref{eq: f average}, and \eqref{eq: fg average} are convenient forms for computing Fermi surface--weighted integrals.

Next, let us compute the inner product $\innera{\phi_\alpha}{\phi_\beta}_\epsilon$, which, as we show in Subsec.~\ref{subsec: GL}, appears in the quadratic term in the Ginzburg-Landau free energy. To linear order in strain, the rescaled functions are
\begin{align}
   \tilde{\phi}_\alpha(\vq)\equiv \phi_\alpha(S^{-1}\vq)\simeq \frac{\sqrt{2}}{k_{\text{F}}}\sum_\beta \left(\delta_{\alpha\beta} - \frac{\delta M^{-1}_{\alpha\beta}}{2m_0^{-1}} \right)q_\beta.
\end{align}
Using Eq.~\eqref{eq: fg average}, we obtain
\begin{equation}\label{eq: inner phi2}
	\innera{\phi_\alpha}{\phi_\beta}_{\epsilon}\simeq \delta_{\alpha\beta} - \frac{\delta M^{-1}_{\alpha\beta}}{ m_0^{-1}}.
\end{equation}
In the absence of strain, $\phi_x$ and $\phi_y$ are orthogonal because they belong to the two-dimensional $E_u$ irrep of $D_{4h}$. Once strain is applied, they transform as independent copies of the same ($B_u$) representation of $C_{2h}$, and their orthogonality is lost.

\subsection{Ginzburg-Landau free energy}
\label{subsec: GL}
To derive the Ginzburg–Landau free energy, we introduce a bosonic order-parameter field by a Hubbard-Stratonovich transformation and integrate out the fermionic degrees of freedom. The partition function can be expressed in terms of a functional integral as $Z = \int \mathcal{D}(\bar{c},c) e^{-S[\bar{c},c]}$, with action
\begin{equation}
    S[\bar{c},c]= \int_0^\beta \d \tau \left[\sum_{\vk}\bar{c}_{\vk}(\tau)(\partial_\tau + \xi_\vk)c_\vk(\tau) -\frac{1}{2\mathcal{V}}\sum_{\vk\vkp\vq}V_{\vk\vkp}\bar{c}_{\vk_+}(\tau)\bar{c}_{-\vk_-}(\tau)c_{-\vkp_-}(\tau)c_{\vkp_+}(\tau) \right].
\end{equation}
Here, $\bar{c}_\vk$ and $c_\vk$ are Grassmann fields associated with the fermionic operators $\hat{c}^\dagger_\vk$ and $\hat{c}_\vk$, respectively. To obtain an effective description of the superconducting phase transition, we decouple the four-fermion interaction in the Cooper channel by introducing bosonic auxiliary fields $\bar{\Delta}_\alpha(\tau,\vq)$ and $\Delta_\alpha(\tau,\vq)$ through the Gaussian path-integral identity
\begin{align}
&\expb{
   \frac{1}{2\mathcal{V}}\int \d \tau \sum_{\vk\vkp\vq} V_{\vk\vkp}\bar{c}_{\vk_+}\bar{c}_{-\vk_-}c_{-\vkp_-}c_{\vkp_+}
}\nonumber\\
&= \int \mathcal{D}(\bar{\Delta},\Delta)
\expb{-\frac{1}{2\mathcal{V}}\int \d \tau \sum_{\vq,\alpha}
\left(
   \frac{|\Delta_\alpha(\vq)|^2}{V}
   - \Delta_\alpha(\vq)\sum_\vk \phi_\alpha(\vk)\bar{c}_{\vk_+}\bar{c}_{-\vk_-}
   - \bar{\Delta}_\alpha(\vq)\sum_\vk \phi_\alpha^*(\vk)c_{-\vk_-}c_{\vk_+}
\right)}.
\end{align}
Furthermore, we switch to the Matsubara representation using the Fourier decomposition
\begin{align}
    \bar{c}_{\vk}(\tau)&= \frac{1}{\sqrt{\beta}}\sum_{ik_n}e^{ik_n\tau}\bar{c}_k,&c_{\vk}(\tau)&= \frac{1}{\sqrt{\beta}}\sum_{ik_n}e^{-ik_n\tau}c_k,\\
    \bar{\Delta}_{\alpha}(\tau,\vq)&= \frac{1}{\sqrt{\beta}}\sum_{iq_m}e^{iq_m\tau}\bar{\Delta}_\alpha(q),&\Delta_{\vk}(\tau,\vq)&= \frac{1}{\sqrt{\beta}}\sum_{iq_m}e^{-iq_m\tau}\Delta_\alpha(q),
\end{align}
where $k=(ik_n,\vk)^T$ and $q=(iq_m,\vq)^T$, with fermionic and bosonic Matsubara frequencies $k_n$ and $q_m$, respectively. The partition function then becomes $Z = \int \mathcal{D}(\bar{c},c)\mathcal{D}(\bar{\Delta},\Delta) e^{-S[\bar{c},c,\bar{\Delta},\Delta]}$, with action
\begin{equation}
    S[\bar{c},c,\bar{\Delta},\Delta]=   \frac{1}{2\mathcal{V}}\sum_q \sum_\alpha\frac{\bar{\Delta}_\alpha(q)\Delta_\alpha(q)}{V} - \frac{1}{2}\sum_{k,k^\prime}\bar{\Psi}_{k}\mathcal{G}^{-1}(k,k^\prime) \Psi_{k^\prime}.
\end{equation}
Here $\Psi_{k} = (c_k, \bar{c}_{-k})^T$ is the two-component Nambu spinor and 
\begin{align}
    \mathcal{G}^{-1}(k,k^\prime)= \mathcal{G}^{-1}_0(k,k^\prime) - \Sigma(k,k^\prime)
\end{align}
is the inverse Nambu-Gor'kov Green's function, where
\begin{align}
   \mathcal{G}^{-1}_0(k,k^\prime)&\equiv  \delta_{k,k^\prime}\mathcal{G}^{-1}_0(k)=\delta_{k,k^\prime}\begin{bmatrix}
        ik_n - \xi_\vk&\\
        &ik_n+\xi_{\vk}
    \end{bmatrix},\\
    \Sigma\left(k+\frac{q}{2},k-\frac{q}{2}\right) &= \frac{1}{\sqrt{\beta}\mathcal{V}}\begin{bmatrix}
        &-\Delta_\vk(q)\\
        -\bar{\Delta}_{\vk}(-q)&
    \end{bmatrix},
\end{align}
with $\bar{\Delta}_\vk(q)=\sum_\alpha \bar{\Delta}_\alpha(q)\phi_\alpha^*(\vk)$ and $\Delta_\vk(q) = \sum_\alpha \Delta_\alpha(q)\phi_\alpha(\vk)$. Performing the Gaussian path integral over the fermionic fields gives $Z = \int \mathcal{D} (\bar{\Delta},\Delta)e^{-S[\bar{\Delta},\Delta]}$, with effective action
\begin{equation}\label{eq: tr-ln action}
    S[\bar{\Delta},\Delta]=\frac{1}{2\mathcal{V}}\sum_q \sum_\alpha \frac{\bar{\Delta}_\alpha(q)\Delta_\alpha(q)}{V}-\frac{1}{2}\Tr \ln \mathcal{G}^{-1},
\end{equation}
where $\Tr$ denotes a trace over both Nambu indices and frequency–momentum space.

The action in the form Eq.~\eqref{eq: tr-ln action} is amenable to a Ginzburg–Landau expansion in powers of the order parameter field and its derivatives. Supposing $\Delta_\alpha$ is small, we can expand the trace-log term in powers of $\Delta_\alpha$ as
\begin{align}
    \Tr \ln \mathcal{G}^{-1} &= \Tr \ln (\mathcal{G}_0^{-1}- \Sigma)= \Tr \ln \mathcal{G}_0^{-1}(1- \mathcal{G}_0 \Sigma)= \Tr \ln \mathcal{G}^{-1}_0 + \Tr \ln (1-\mathcal{G}_0 \Sigma)\nonumber\\
    &= \Tr \ln \mathcal{G}^{-1}_0 - \sum_{m=1}^\infty T^{(m)}\label{eq: tr-ln expansion}.
\end{align}
The first term is independent of $\Delta_\alpha$ and may be dropped. The $m$th order term is
\begin{equation}
    T^{(m)}\equiv \frac{\Tr(\mathcal{G}_0\Sigma)^m}{m}.
\end{equation}
Note that because of the matrix structure of $\mathcal{G}_0$ and $\Sigma$ in Nambu space, the only nonzero terms are of even order.

The leading contribution arises at quadratic order in $\Delta_\alpha$. It can be written as
\begin{align}
    T^{(2)}&=\frac{1}{2} \Tr \mathcal{G}_0 \Sigma \mathcal{G}_0 \Sigma \nonumber\\
    &= \frac{1}{2}\sum_{k,q}\tr \mathcal{G}_0\left(k+\frac{q}{2}\right)\Sigma\left(k+\frac{q}{2},k-\frac{q}{2}\right)\mathcal{G}_0\left(k-\frac{q}{2}\right)\Sigma\left(k-\frac{q}{2},k+\frac{q}{2}\right)\nonumber \\
    &= \frac{1}{\beta \mathcal{V}^2}\sum_{k,q}\frac{\bar{\Delta}_\vk(q)\Delta_\vk(q)}{(ik_n +iq_m/2- \xi_{\vk_+})(ik_n- i q_m/2 +\xi_{\vk_-})}.
\end{align}
In the second line, $\tr$ denotes a trace over the Nambu indices only. Expanding to zeroth order in $iq_m$ (static limit) and second order in $\vq$, and evaluating the Matsubara sum, we obtain
\begin{align}
    \frac{1}{\beta}\sum_{ik_n}\frac{1}{(ik_n+iq_m/2-\xi_{\vk_+})(ik_n-iq_m/2+\xi_{\vk_-})}&\simeq \frac{1}{\beta}\sum_{ik_n}\left[\frac{1}{(ik_n)^2-\xi_\vk^2}+\sum_{ij}\frac{\hbar^2 v_i v_j q_i q_j}{4}\frac{3(ik_n)^2 + \xi_\vk^2}{[(ik_n)^2-\xi_\vk^2]^3}\right]\nonumber\\
    &=-\left[\frac{1-2n_{\text{F}}(\xi_\vk)}{2\xi_\vk} - \sum_{ij}\frac{\hbar^2 v_i v_j q_i q_j}{4}\frac{n_{\text{F}}^{\prime\prime}(\xi_\vk)}{2\xi_\vk}\right],
\end{align}
where $n_{\text{F}}(\xi)=(e^{\beta\xi}+1)^{-1}$ is the Fermi–Dirac distribution. In the expansion, we have omitted terms linear in the Fermi velocity, which vanish once the angular integral over $\vk$ is performed, and terms odd in $\xi_\vk$, which vanish once the radial integral is performed. Therefore, 
\begin{align}
    T^{(2)}&=-\frac{1}{\mathcal{V}^2}\sum_{q}\sum_\vk\bar{\Delta}_\vk(q) \left[\frac{1-2n_{\text{F}}(\xi_\vk)}{2\xi_\vk} - \sum_{ij}\frac{\hbar^2 v_i v_j q_i q_j}{4}\frac{n^{\prime\prime}_{\text{F}}(\xi_\vk)}{2\xi_\vk}\right]\Delta_\vk(q)\nonumber\\
    &=-\frac{1}{\mathcal{V}}\sum_q\left[ N_\epsilon(0)\int \d \xi \frac{1-2n_{\text{F}}(\xi)}{2\xi}\expect{|\Delta_\vk(\vq)|^2}_{\epsilon}- \sum_{ij}\frac{\hbar^2q_iq_j}{4}N_\epsilon(0)\int \d \xi \frac{n_{{\text{F}}}^{\prime\prime}(\xi)}{2\xi} \expect{v_iv_j |\Delta_\vk(\vq)|^2}_\epsilon\right] \nonumber\\
    &=- \frac{1}{\mathcal{V}}\sum_q \sum_{\alpha\beta}\bar{\Delta}_\alpha(q)\left[N_\epsilon(0)\ln \left(\frac{2e^\gamma}{\pi}\frac{\Lambda}{k_BT} \right)\innera{\phi_\alpha}{\phi_\beta}_{\epsilon}- \sum_{ij}\frac{7\zeta(3)}{16\pi^2}\frac{N_\epsilon(0)}{(k_BT)^2} \hbar^2 q_iq_j\expect{v_iv_j\phi_\alpha^*\phi_\beta}_{\epsilon}\right]\Delta_\beta(q),\label{eq: T2}
\end{align}
where $\gamma$ is the Euler–Mascheroni constant and $\xi(z)$ is the Riemann zeta function.

The next nonvanishing contribution arises at quartic order. To zeroth order in $q$, it is given by
\begin{align}
    T^{(4)}&\simeq \frac{1}{2\beta^2 \mathcal{V}^4}\sum_k\sum_{q_1,q_2,q_3}\frac{\bar{\Delta}_\vk(q_1)\bar{\Delta}_\vk(q_2)\Delta_\vk(q_3)\Delta_\vk(q_1+q_2-q_3)}{(ik_n-\xi_\vk)^2(ik_n+\xi_\vk)^2}.
\end{align}
Evaluating the Matsubara sum and performing the radial integral,
\begin{align}
    T^{(4)}&\simeq  \frac{N_\epsilon(0)k_B T}{2}\frac{1}{\mathcal{V}^3}\sum_{q_1q_2q_3}\sum_{\alpha\beta\gamma\delta} \int\d\xi \left[ \frac{n_{\text{F}}^\prime(\xi_\vk)}{2\xi_\vk^2}+\frac{1-2n_{\text{F}}(\xi_\vk)}{4\xi_\vk^3}\right]\innera{\phi_\alpha\phi_\beta}{\phi_\gamma\phi_\delta}_\epsilon\bar{\Delta}_\alpha(q_1)\bar{\Delta}_\beta(q_2)\Delta_\gamma(q_3)\Delta_\delta(q_1+q_2-q_3) \nonumber\\
    &=\frac{7\zeta(3)}{16\pi^2}\frac{N_\epsilon(0)}{k_BT} \frac{1}{\mathcal{V}^3}\sum_{q_1q_2q_3}\sum_{\alpha\beta\gamma\delta}\innera{\phi_\alpha\phi_\beta}{\phi_\gamma\phi_\delta}_\epsilon\bar{\Delta}_\alpha(q_1)\bar{\Delta}_\beta(q_2)\Delta_\gamma(q_3)\Delta_\delta(q_1+q_2-q_3).\label{eq: T4}
\end{align}

Combining Eqs.~\eqref{eq: tr-ln action}, \eqref{eq: tr-ln expansion}, \eqref{eq: T2}, and \eqref{eq: T4}, and taking the Fourier transform back to real space and imaginary time, we obtain
\begin{equation}
    S[\bar{\Delta},\Delta]= \int \d \tau\, \d^3 r \left[ \sum_{\alpha\beta}\bar{\Delta}_\alpha A_{\alpha\beta}\Delta_\beta + \sum_{ij,\alpha\beta}K_{ij,\alpha\beta}\partial_i \bar{\Delta}_\alpha \partial_j \Delta_\beta+ \frac{1}{2}\sum_{\alpha\beta\gamma\delta}B_{\alpha\beta\gamma\delta}\bar{\Delta}_\alpha\bar{\Delta}_\beta \Delta_\gamma \Delta_\delta\right],
\end{equation}
in which the coefficients are
\begin{align}
    A_{\alpha\beta}&=\frac{\delta_{\alpha\beta}}{2V}-\frac{N_\epsilon(0)}{2}\ln \left( \frac{2e^\gamma}{\pi}\frac{\Lambda}{k_BT}\right)\innera{\phi_\alpha}{\phi_\beta}_\epsilon,\\
    K_{ij,\alpha\beta}&=\frac{7\zeta(3)}{32\pi^2}\frac{N_\epsilon(0)}{(k_BT)^2} \hbar^2 \expect{v_i v_j\phi_\alpha^*\phi_\beta}_{\epsilon},\\
    B_{\alpha\beta\gamma\delta}&=\frac{7\zeta(3)}{16\pi^2}\frac{N_\epsilon(0)}{(k_BT)^2}\innera{\phi_\alpha\phi_\beta}{\phi_\gamma\phi_\delta}_\epsilon.
\end{align}
In the mean-field approximation, $\Delta_\alpha$ does not vary in imaginary time, and we obtain the free energy $F[\bar{\Delta},\Delta]\equiv S[\bar{\Delta},\Delta]/\beta$ given in the main text. In the absence of strain, the two partial wave channels are degenerate with critical temperature $T_c^0=2\Lambda e^\gamma e^{-1/[N_0(0)V]}/\pi $. Expanding the quadratic term for small strain and temperature deviation from $T_c^0$, we obtain
\begin{align}
    A_{\alpha\beta}\simeq \frac{\delta_{\alpha\beta}}{2}\left[N_0(0) t - \frac{(C-2)\epsilon_0}{V} \right]+ \frac{1}{2V}\frac{\delta M^{-1}_{\alpha\beta}}{m_0^{-1}},
\end{align}
where $t=(T-T_c^0)/T_c^0$.

\section{D\lowercase{erivation of the upper critical field}}
In this SM, we provide a detailed derivation of the upper critical field $B_{c2}$ for a magnetic field in the $x$-$y$ plane when the superfluid stiffness tensor is anisotropic [Eq.~\eqref{eq: Bc2} of the main text]. 

We begin with the equation of motion associated with the Ginzburg-Landau free energy [Eq.~\eqref{eq: projected F} of the main text]:
\begin{equation}
    0= -\sum_{jk}K_{jk}\partial_j \partial_k\psi(\vr) + a_- \psi(\vr) + b_- |\psi(\vr)|^2 \psi(\vr).
\end{equation}
In the vicinity of the phase transition, $\psi(\vr)$ is small and thus the cubic term can be neglected. We apply a uniform in-plane magnetic field $\vec{B}= B(\cos\theta,\sin\theta,0)^T$ to the system, which is incorporated by minimal substitution:
\begin{equation}
    0= \frac{1}{\hbar^2}\sum_{jk}K_{jk}\left(-i\hbar\partial_j- e^*A_j\right)\left(-i\hbar\partial_k - e^*A_k \right)\psi(\vr) + a_- \psi(\vr),
\end{equation}
where $e^*=2e$ is the charge of the Cooper pair. Here, we adopt the Landau-like gauge $\vec{A}(\vr)=B(0,0,y\cos\theta- x\sin\theta)^T$.

The magnetic field renormalizes the coefficient $a_-$, thereby shifting the critical temperature. As with the vortex problem in the main text, we first perform the change of coordinates $\vrp= \xi^{-1}\vr$, where $\xi=\sqrt{K}/\sqrt{|a_-|}$, to bring the Ginzburg-Landau equation to an isotropic form. Under this transformation, the derivative transforms as
\begin{equation}
    \partial_i^\prime = \sum_j (\partial_i^\prime r_j) \partial_j =\sum_j \xi_{ji} \partial_j.
\end{equation}
Using this identity, the Ginzburg-Landau equation becomes
\begin{align}
	0&= \left[\frac{1}{\hbar^2}\sum_{jk}K_{jk}\left(-i\hbar\partial_j -e^*A_j \right)\left(-i\hbar\partial_k - e^*A_k \right)+a_-\right]\psi(\vr)\nonumber\\
    &=\left[\frac{|a_-|}{\hbar^2}\sum_{jkl}\xi_{jl}\xi_{lk}(-i\hbar\partial_j - e^*A_j)(-i\hbar\partial_k - e^*A_k)+a_-\right]\psi(\vr)\nonumber\\
	&=\left[\frac{|a_-|}{\hbar^2}\sum_l\left(-i\hbar\partial_l^\prime - e^*A_l^\prime \right)\left(-i\hbar\partial_l^\prime - e^*A_l^\prime \right)+a_-\right]\psi(\vrp),
\end{align}
where $A^\prime_i=\sum_j \xi_{ij}A_{j}$. In reaching the third line, we have used that $\xi$ is symmetric. The first term is the Schr\"odinger Hamiltonian for a non-relativistic particle in a magnetic field $\vec{B}^\prime(\vrp)=\nabla^\prime \times \vec{A}^\prime(\vrp)$. As we show below, the vector potential $\vec{A}^\prime(\vrp)$ preserves translational symmetry in the $z^\prime$ direction and corresponds still to a uniform in-plane magnetic field $\vec{B}^\prime(\vrp)$. Thus, we look for solutions of the form $\psi(\vrp)= \psi_0 \Psi_{k_z^\prime,n}(\vrp)$, where $\Psi_{k_z^\prime,n}(\vrp)$ are the normalized Landau level eigenfunctions satisfying
\begin{equation}
    \frac{1}{\hbar^2}\sum_l \left(-i\hbar\partial_l^\prime - e^*A_l^\prime \right) \left(-i\hbar\partial_l^\prime - e^*A_l^\prime \right) \Psi_{k_z,n}(\vrp) = \left[ k_z^{\prime 2}+\frac{2e^*B^\prime}{\hbar}\left(n+\frac{1}{2} \right)\right]\Psi_{k_z,n}(\vrp),
\end{equation}
and $n\in \mathbb{Z}_{\geq 0}$. Thus, the Ginzburg-Landau equation becomes
\begin{equation}
    0= \left[|a_-|\left(k_z^{\prime 2} +\frac{2e^*B^\prime}{\hbar}\left(n+\frac{1}{2} \right)\right)+a_-\right]\psi_0.
\end{equation}
The phase transition occurs when the coefficient in the linear term becomes zero. Clearly, this can only happen when $a_-<0$, and it first happens for $k_z^\prime=0$ and $n=0$. This gives the critical condition
\begin{equation}\label{eq: Bprime}
    B^\prime(B_c) = \frac{\Phi_0}{2\pi},
\end{equation}
where $\Phi_0=h/e^*$ is the superconducting flux quantum.

We now determine the effective magnetic field $\vec{B}^\prime(\vrp)=\nabla^\prime \times \vec{A}^\prime(\vrp)$. First, we need to express the effective vector potential $\vec{A}^\prime(\vrp)=\xi \vec{A}(\vr(\vrp))$ as a function of $\vrp$. By explicit computation,
\begin{align} 
	\vec{A}^\prime(\vrp)&= \xi \vec{A}(\vr(\vrp)) \nonumber\\
    &=B\xi_{zz}\left(y\cos\theta-x\sin\theta \right)\hat{\vec{z}}\nonumber\\
	&=B \xi_{zz}\begin{pmatrix}
		-\sin\theta&\cos\theta
	\end{pmatrix}
\begin{pmatrix}
	x\\
	y
\end{pmatrix}\hat{\vec{z}}\nonumber\\
&=B \xi_{zz}\begin{pmatrix}
	-\sin\theta&\cos\theta
\end{pmatrix}
\begin{pmatrix}
	\xi_{xx}&\xi_{xy}\\
	\xi_{xy}&\xi_{yy}
\end{pmatrix}
\begin{pmatrix}
	x^\prime\\
	y^\prime
\end{pmatrix}\hat{\vec{z}}\nonumber\\
&=B_{zz}\xi_{zz}\begin{pmatrix}
	-\sin\theta\xi_{xx}+\cos\theta \xi_{xy}&-\sin\theta\xi_{xy}+\cos\theta\xi_{yy}
\end{pmatrix}
\begin{pmatrix}
	x^\prime\\
	y^\prime
\end{pmatrix}\hat{\vec{z}}.
\end{align}
Since $\vec{A}^\prime$ does not depend on $z^\prime$, it preserves translational symmetry in that direction, as asserted above. Taking the curl, we obtain the effective magnetic field
\begin{align}
	\vec{B}^\prime(\vrp)&\equiv \nabla^\prime \times \vec{A}(\vrp)\nonumber\\
    &=B\xi_{zz}\begin{pmatrix}
	-\sin\theta\xi_{xy}+\cos\theta\xi_{yy}\\
		\sin\theta\xi_{xx}-\cos\theta \xi_{xy}\\
		0
	\end{pmatrix}\nonumber\\
&=B\xi_{zz}\begin{pmatrix}
	\xi_{yy}&-\xi_{xy}&0\\
	-\xi_{xy}&\xi_{xx}&0\\
    0&0&\xi_{zz}
\end{pmatrix}
\begin{pmatrix}
	\cos\theta\\
	\sin\theta\\
    0
\end{pmatrix}.
\end{align}
To compute the magnitude of the effective magnetic field, first note that we can write
\begin{equation}
    \begin{pmatrix}
        \xi_{yy}&-\xi_{xy}&0\\
        -\xi_{xy}&\xi_{xx}&0\\
        0&0&\xi_{zz}
    \end{pmatrix}
    =\underbrace{\begin{pmatrix}
        0&1&0\\
        -1&0&0\\
        0&0&1
    \end{pmatrix}}_{R^T}
    \begin{pmatrix}
        \xi_{xx}&\xi_{xy}&0\\
        \xi_{xy}&\xi_{yy}&0\\
        0&0&\xi_{zz}
    \end{pmatrix}
    \underbrace{\begin{pmatrix}
        0&-1&0\\
        1&0&0\\
        0&0&1
    \end{pmatrix}}_R,
\end{equation}
where $R$ is a $\pi/2$ rotation about the $\hat{\vz}$ axis. Hence,
\begin{align}
    \vec{B}^\prime&=B \xi_{zz}R^T \xi R \hat{\vec{n}}\nonumber\\
    &=B \xi_{zz}R^T \xi \hat{\vec{n}}_\perp,
\end{align}
where $\hat{\vec{n}}_\perp = R\hat{\vec{n}}=\hat{\vec{z}}\times \hat{\vec{n}}$. This corresponds to a uniform field, as claimed above, with magnitude
\begin{align}
	B^\prime = B \xi_{zz} \xi_\perp(\hat{\vec{n}}),
\end{align}
where $\xi_\perp(\hat{\vec{n}})=\sqrt{\hat{\vec{n}}_\perp \xi^2 \hat{\vec{n}}_\perp}$. Substituting this into Eq.~\eqref{eq: Bprime}, we obtain
\begin{equation}
    B_c = \frac{\Phi_0}{2\pi\xi_{zz}\xi_\perp(\hat{\vec{n}})},
\end{equation}
as stated in the main text.

\section{W\lowercase{inding of the thermal conductivity tensor}}
In this SM, we provide further details regarding the winding number of the thermal conductivity tensor $\kappa_{\alpha\beta}$. As mentioned in the main text, the components of the thermal conductivity tensor $\kappa_3=(\kappa_{xx}-\kappa_{yy})/2$ and $\kappa_1=\kappa_{xy}$ exhibit a similar winding behavior as the superfluid stiffness tensor. However, as we explain below, the origin of this winding comes from a combination of gap anisotropy and Fermi surface anisotropy, unlike the superfluid stiffness tensor, which to lowest order is due solely to the gap anisotropy.

The in-plane thermal conductivity tensor in the superconducting state within the Born approximation is given by~\cite{Mineev1999Introduction}
\begin{equation}\label{eq: kappa}
    \kappa_{\alpha\beta}= \frac{N_\epsilon(0)\tau}{T}\int \frac{\d^3k}{(2\pi)^3} E_\vk^2 v^{\text{s}}_{\vk,\alpha}v^{\text{s}}_{\vk,\beta}\left[-\partiald{n_{\text{F}}(E_\vk)}{E_\vk} \right]\frac{1}{N^{\text{s}}_\epsilon(E_\vk)},
\end{equation}
where $\tau$ is the normal state scattering time, $E_\vk = \sqrt{\xi_\vk^2 + |\Delta_\vk|^2}$ is the quasiparticle dispersion, $\vv_\vk^{\text{s}} = \nabla_\vk E_\vk/\hbar$ is the quasiparticle velocity, and $N_{\epsilon}^{\text{s}}(E_\vk)$ is the density of states in the superconducting state. We are primarily interested in the components $\kappa_3$ and $\kappa_1$, which both vanish in the presence of tetragonal symmetry. There are two sources of anisotropy for the thermal conductivity tensor: the anisotropy in the Fermi surface, which is present in both the normal and superconducting states, and that of the gap, present only in the superconducting state. In the two subsections below, we compute the in-plane thermal conductivity in the normal ($\kappa_{\alpha\beta}^{\text{n}}$) and superconducting ($\kappa_{\alpha\beta}^{\text{s}}$) states, highlighting the origin of the anisotropy in each case.

\subsection{Normal state}
Let us first consider the normal state thermal conductivity in the presence of strain. Taking the limit $\Delta_\vk\to 0$, Eq.~\eqref{eq: kappa} simplifies to
\begin{equation}
    \kappa^{\text{n}}_{\alpha\beta}(\epsilon)\simeq \frac{\tau}{T_c^0} \int \frac{\d^3k}{(2\pi)^3}\xi_\vk^2(\epsilon) v_{\vk,\alpha}^{\text{n}}(\epsilon)v^{\text{n}}_{\vk,\beta}(\epsilon)\left[ -\partiald{n_{\text{F}}(\xi_\vk)}{\xi_\vk}\right],
\end{equation}
where $v^{\text{n}}_{\vk}(\epsilon)=\nabla_\vk \xi_\vk(\epsilon)/\hbar$ is the normal state Fermi velocity in the presence of strain. 
Using Eq.~\eqref{eq: I2} to evaluate the integral, we have
\begin{align}
    \kappa^{\text{n}}_{\alpha\beta}(\epsilon)= \frac{\tau N_\epsilon(0)}{T_c^0}\underbrace{\int_{-\Lambda}^{\Lambda} \d\tilde{\xi}\,\tilde{\xi}^2\left[-\partiald{n_{\text{F}}(\tilde{\xi})}{\tilde{\xi}} \right]}_{\frac{\pi^2}{3}(k_BT_c^0)^2} \expect{v^{\text{n}}_\alpha(\epsilon) v^{\text{n}}_\beta(\epsilon)}_\epsilon.
\end{align}
In other words, the normal state thermal conductivity is proportional to the velocity-velocity correlator $\expect{v^{\text{n}}_\alpha(\epsilon) v^{\text{n}}_\beta(\epsilon)}_\epsilon$. In the presence of tetragonal symmetry, it is proportional to the identity. However, under strain, the components $\kappa^{\text{n}}_3\equiv (\kappa^{\text{n}}_{xx}-\kappa^{\text{n}}_{yy})/2$ and $\kappa^{\text{n}}_1\equiv \kappa^{\text{n}}_{xy}$ can become nonzero. The normal state Fermi velocity is
\begin{equation}
    v^{\text{n}}_{\vq,\alpha}(\epsilon)\equiv \frac{1}{\hbar}\partial_{q_\alpha}\xi_\vq(\epsilon)=\hbar \sum_{\beta}M^{-1}_{\alpha\beta}q_\beta.
\end{equation}
Hence,
\begin{equation}
    \tilde{v}^{\text{n}}_{\vq,\alpha}\equiv v^{\text{n}}_{S^{-1}\vq,\alpha}\simeq \frac{\hbar}{m_0}\sum_\beta\left(\delta_{\alpha\beta}+\frac{\delta M^{-1}_{\alpha\beta}}{2m_0^{-1}} \right)q_\beta.
\end{equation}
The Fermi surface average is then
\begin{align}
    \expect{v^{\text{n}}_\alpha(\epsilon) v^{\text{n}}_\beta(\epsilon)}_\epsilon&\simeq  \frac{v_{\text{F}}^2}{2}\left( \delta_{\alpha\beta}+  \frac{\delta M^{-1}_{\alpha\beta}}{m_0^{-1}}\right),
\end{align}
where $v_{\text{F}}=\hbar k_{\text{F}}/m_0$. Therefore, the normal state conductivity is
\begin{equation}\label{eq: normal state kappa}
    \frac{\kappa^{\text{n}}_{\alpha\beta}(\epsilon)}{\kappa^{\text{n}}_0(0)} = \frac{N_\epsilon(0)}{N_0(0)}\left(\delta_{\alpha\beta} + \frac{\delta M^{-1}_{\alpha\beta}}{m_0^{-1}}\right),
\end{equation}
where $\kappa_0^{\text{n}}(0)=\tau N_0(0)v_{\text{F}}^2\pi^2k_B^2 T_c^0/6$ is the normal state thermal conductivity in the absence of strain.

\begin{figure}[t!]
\centering
\includegraphics[width=0.4\columnwidth]{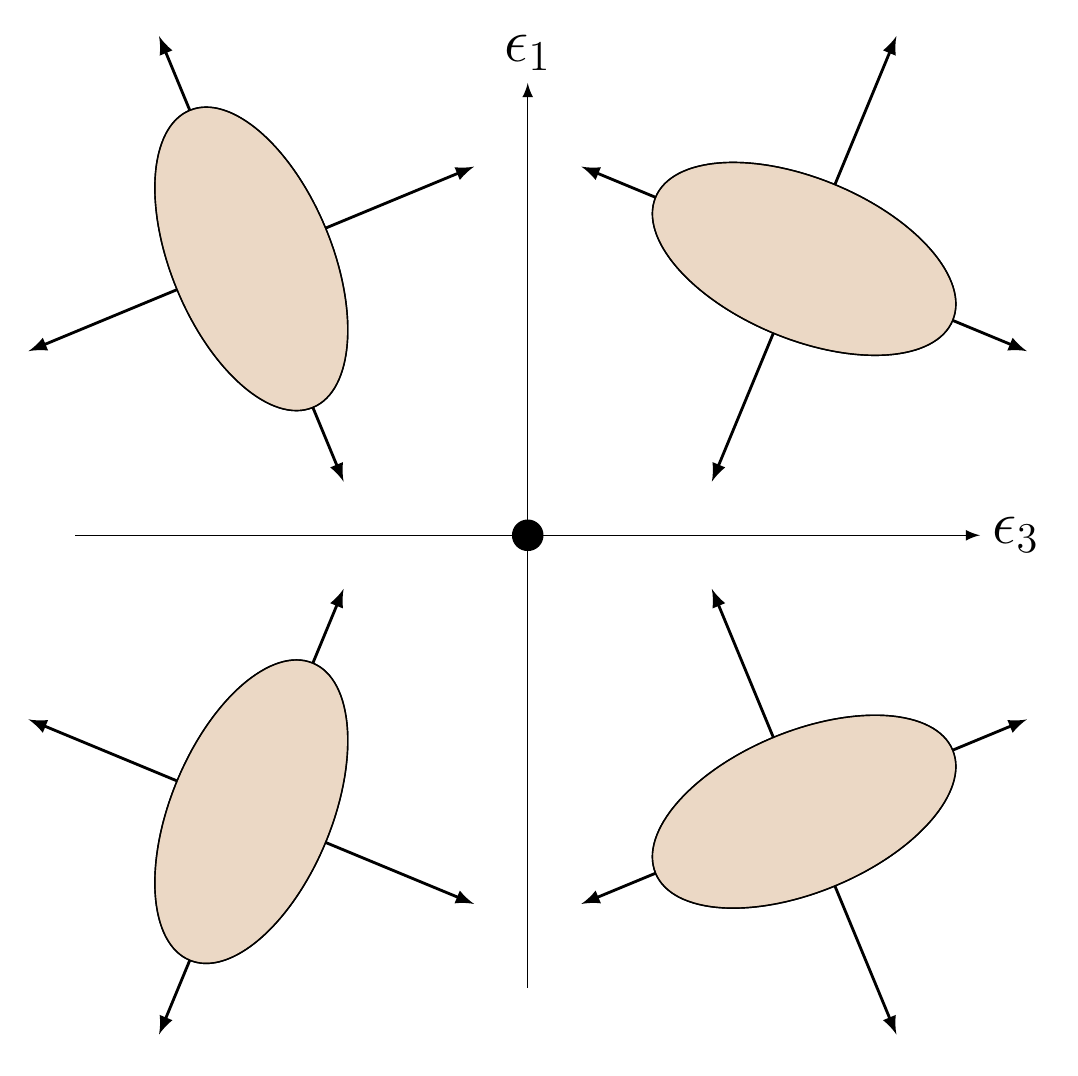}
\caption{\label{fig: FS} The Fermi surface as a function of strain. Under strain, the Fermi surface is elliptical (peach disk). The Fermi velocities along its semimajor and semiminor axes are shown as black arrows. The speed is greater along the semiminor axes than along the semimajor axes.}
\end{figure}

The coefficients $\kappa_3^{\text{n}}$ and $\kappa_1^{\text{n}}$ exhibit a similar winding to $K_3$ and $K_1$, except they are out of phase by $\pi$. From Eq.~\eqref{eq: normal state kappa}, we can immediately read off
\begin{align}
    \frac{\kappa^{\text{n}}_3(\epsilon)}{\kappa^{\text{n}}_0(0)}&=-\frac{N_\epsilon(0)}{N_0(0)}\cos\varphi,&\frac{\kappa^{\text{n}}_1(\epsilon)}{\kappa^{\text{n}}_0(0)}&=\frac{N_\epsilon(0)}{N_0(0)}\sin\varphi.
\end{align}
The winding can be understood as follows. The strain changes the velocity-velocity correlator in two ways. First, it makes the Fermi velocity anisotropic: Under strain, the Fermi surface becomes elliptical, and the speed at the semiminor axes is greater than that at the semimajor axes (Fig.~\ref{fig: FS}). Second, it alters the distribution of states around the Fermi surface. In particular, it reduces the density of states around the semiminor axes and enhances it around the semimajor axes. These two effects compete in the Fermi surface average. However, the anisotropy of the Fermi velocity dominates over that of the density of states. Therefore, $\kappa_3^{\text{n}}$ and $\kappa_1^{\text{n}}$ probe the direction of the semiminor axes, which coincide with the nodal directions of the superconducting state (Fig.~\ref{fig: orbitals}). In contrast, $K_3$ and $K_1$ probe the gap maxima, located along the semimajor axes. Since the semimajor and semiminor axes are related by a $\pi/2$ rotation, and such a rotation corresponds to a $\pi$ shift in $\varphi$, this accounts for the observed $\pi$ phase difference.

\subsection{Superconducting state}
We now turn to the superconducting case. We assume that, at low temperatures, the system remains in the intermediate state. Because of Fermi statistics, the dominant contributions arise from low-energy quasiparticle states near the nodes. Consequently, $\kappa^{\text{s}}_3$ and $\kappa^{\text{s}}_1$ are expected to exhibit the same winding behavior as $\kappa^{\text{n}}_3$ and $\kappa^{\text{n}}_1$. Since the Fermi surface anisotropy enhances the anisotropy of the gap function and would not change the winding, for simplicity, we perform the Fermi surface integral over the unstrained Fermi surface.

We begin from Eq.~\eqref{eq: kappa},
\begin{equation}
    \kappa^{\text{s}}_{\alpha\beta}= \frac{N_0(0)\tau}{T}\int \frac{\d^3k}{(2\pi)^3} E_\vk^2 v^{\text{s}}_{\vk,\alpha}v^{\text{s}}_{\vk,\beta}\left[-\partiald{n_{\text{F}}(E_\vk)}{E_\vk} \right]\frac{1}{N^{\text{s}}_0(E_\vk)}.
\end{equation}
Let us now bring the thermal conductivity to a more convenient form. The quasiparticle velocity is given by
\begin{equation}
	v_{\vk,\alpha} = \frac{\xi_\vk}{E_\vk}\frac{1}{\hbar}\partial_{k_\alpha} \xi_\vk + \mathcal{O}\left(\frac{\Delta_\vk}{E_{\text{F}}}\right)\simeq  \frac{\xi_\vk}{E_\vk} v_{\text{F}}\frac{k_\alpha}{k_{\text{F}}}.
\end{equation}
Performing the change of coordinates from $\xi_\vk$ to $E_{\vk}= \sqrt{\xi_\vk^2 + |\Delta_\vk|^2}$, the thermal conductivity becomes
\begin{align}\label{eq: kappa1}
	\kappa^{\text{s}}_{\alpha\beta} = \frac{2[N_0(0)]^2v_{\text{F}}^2\tau}{k_BT^2}  \int_0^\Lambda \d E \frac{E}{4\cosh^2(\beta E/2)}\frac{1}{N^{\text{s}}_0(E)}\int \frac{\d\theta_\vk}{2\pi}  	\frac{k_\alpha}{k_{\text{F}}} \frac{k_\beta}{k_{\text{F}}}\sqrt{E^2-|\Delta_\vk|}\theta(E-|\Delta_\vk|).
\end{align}
Normalizing again by $\kappa_0^{\text{n}}(0)$, the in-plane thermal conductivity is given by
\begin{equation}\label{eq: kappa num}
    \frac{\kappa_{\alpha\beta}^{\text{s}}}{\kappa^{\text{n}}_0(0)}=\frac{12N_0(0)}{\pi^2 k_B^3 T^2 T_c^0}     \int_0^\Lambda \d E \frac{E}{4\cosh^2(\beta E/2)}\frac{1}{N^{\text{s}}_0(E)}\int \frac{\d\theta_\vk}{2\pi}  	\frac{k_\alpha}{k_{\text{F}}} \frac{k_{\beta}}{k_{\text{F}}}\sqrt{E^2-|\Delta_\vk|^2}\theta(E-|\Delta_\vk|).
\end{equation}

To proceed, we need to determine the superconducting density of states, which admits an exact analytic expression for a $p$-wave superconductor. The superconducting density of states is defined as
\begin{equation}
	N_0^{\text{s}}(E) = \frac{1}{2}\int \frac{\d^3k}{(2\pi)^3}\delta(E-E_\vk),\hspace{5mm}E>0.
\end{equation}
The factor of $1/2$ accounts for the fact that only half of the modes in the Brillouin zone are independent. Then, performing a change of variables, 
\begin{align}
	N_0^{\text{s}}(E)& = \frac{N_0(0)}{2}\int_{-\Lambda}^\Lambda \d \xi_\vk \int \frac{\d\theta_\vk}{2\pi}\delta(E-E_\vk)\nonumber \\
	&=N_0(0) \int \frac{\d\theta_\vk}{2\pi}  \frac{E}{\sqrt{E^2 - |\Delta_\vk|^2}}\theta(E - |\Delta_\vk|).\label{eq: Ns1}
\end{align}
If the gap has the form $\Delta_\vk = \sqrt{2}\psi_0\cos(\theta_\vk+\varphi/2)$, where $\psi_0$ is real and positive, this integral admits an analytical solution:
\begin{equation}\label{eq: Ns exact}
    N^{\text{s}}_0(E) = N_0(0)\frac{2}{\pi}\times\begin{cases}
		E K(E^2/(\sqrt{2}\psi_0)^2)/[\sqrt{2}\psi_0],& 0<E< \sqrt{2}\psi_0,\\
		K\left(2\psi_0^2/E^2\right),&E>\sqrt{2}\psi_0,
	\end{cases}
\end{equation}
where $K(m)$ is the complete elliptic integral of the first kind. This reduces to the normal state density of states when the superconducting order parameter $\psi_0\to 0$.

At low temperatures, the dominant contribution in the integral Eq.~\eqref{eq: kappa num} comes from the nodes, which are located at $\theta_\vk = \theta_{\pm}\equiv \pm \pi/2-\varphi/2$. Linearizing about the two nodes $\theta_\pm $ allows us to simplify the angular integral as
\begin{align}
    J_{\alpha\beta}(E) &\equiv \int \frac{\d\theta_\vk}{2\pi}\frac{k_\alpha(\theta_\vk)}{k_{\text{F}}}\frac{k_\beta(\theta_\vk)}{k_{\text{F}}}\sqrt{E^2-|\Delta_\vk|^2}\theta(E-|\Delta_\vk|)\nonumber\\
    &=\sum_{n=\pm } \frac{k_\alpha(\theta_n)}{k_{\text{F}}}\frac{k_\beta(\theta_n)}{k_{\text{F}}}\int\frac{\d\theta_\vk}{2\pi}\sqrt{E^2- [\sqrt{2}\psi_0\cos(\theta_\vk+\varphi/2)]^2}\theta(E-\sqrt{2}\psi_0\cos(\theta_\vk+\varphi/2))\nonumber\\
    &\simeq  \sum_{n=\pm } \frac{k_\alpha(\theta_n)}{k_{\text{F}}}\frac{k_\beta(\theta_n)}{k_{\text{F}}}\int_{ E/\sqrt{2}\psi_0}^{-E/\sqrt{2}\psi_0} \frac{\d\delta \theta_\vk}{2\pi} \sqrt{E^2-(\sqrt{2}\psi_0\delta\theta_\vk)^2}\nonumber\\
    &=\sum_{n=\pm}\frac{k_\alpha(\theta_n)}{k_{\text{F}}}\frac{k_\beta(\theta_n) }{k_{\text{F}}}\frac{E^2}{4\sqrt{2}\psi_0}.
\end{align}
Expanding the density of states at low energies, $N^{\text{s}}_0(E)\simeq N_0(0)E/(\sqrt{2}\psi_0)$, we obtain
\begin{align}
    \frac{\kappa_{\alpha\beta}^{\text{s}}}{\kappa_0^{\text{n}}(0)}&=\frac{3}{\pi^2 k_B^3 T^2 T_c^0}    \underbrace{\int_0^\Lambda \d E \frac{E^2}{4\cosh^2(\beta E/2)}}_{\frac{\pi^2}{6}(k_BT)^3} \sum_{n=\pm } \frac{k_\alpha(\theta_n)}{k_{\text{F}}}\frac{k_\beta(\theta_n)}{k_{\text{F}}}\nonumber\\
    &=  \frac{T}{T_c^0}  \begin{bmatrix}
        \sin^2(\phi/2)&\sin(\phi/2)\cos(\phi/2)\\
        \sin(\phi/2)\cos(\phi/2)&\cos^2(\phi/2)
    \end{bmatrix}_{\alpha\beta}.
\end{align}
Thus, we have $\kappa^{\text{s}}_3\sim -\cos\phi$ and $\kappa^{\text{s}}_1\sim \sin\phi$ as stated.

\end{document}